\RequirePackage[l2tabu, orthodox]{nag}
\RequirePackage{snapshot}

\documentclass[10pt,onecolumn]{extarticle}

\makeatletter
\if@twocolumn
  \usepackage[dvips,letterpaper,top=0.5in, bottom=0.5in, left=0.75in, right=0.5in,includefoot,heightrounded]{geometry}
\else
  \usepackage[dvips,letterpaper,margin=1in,includefoot,heightrounded]{geometry}
\fi

\usepackage{srcltx}

\usepackage[russian,portuges,english]{babel}

\iflanguage{portuges}
    {\newcommand{\keywordname}{Palavras-chaves}}
    {\newcommand{\keywordname}{Keywords}}

\usepackage{amsmath}
\usepackage{amssymb}
\usepackage{amsfonts}

\usepackage{abstract}

\usepackage{graphicx}
\usepackage[usenames,dvipsnames,svgnames,x11names]{xcolor}
\usepackage{subfigure}

\usepackage{booktabs}

\usepackage{setspace}
\usepackage{flushend}

\usepackage{multicol}

\usepackage{cite}

\usepackage{hyperref}
\usepackage[normalem]{ulem}

\usepackage{enumerate}

\usepackage[noend]{algpseudocode}

\usepackage{listings}

\usepackage[short,12hr]{datetime}
       \usepackage{fouriernc}
\makeatletter

\newcommand{\printtitle}{%
\makeatletter
\if@twocolumn

\twocolumn[%
  \maketitle
  \begin{onecolabstract}
    \myabstract
  \end{onecolabstract}
  \begin{center}
    \small
    \textbf{\keywordname}
    \\\medskip
    \mykeywords
  \end{center}
  \bigskip
]
\saythanks
\else
  \maketitle
  \begin{onecolabstract}
    \myabstract
  \end{onecolabstract}
  \begin{center}
    \small
    \textbf{\keywordname}
    \\\medskip
    \mykeywords
  \end{center}
  \bigskip
  \onehalfspacing
\fi
\makeatother
}

\author{%
F. M. Bayer%
\thanks{%
Departamento de Estat\'istica and LACESM, Universidade Federal de Santa Maria, Brazil.
E-mail: \url{bayer@ufsm.br}}
\and
A. J. Kozakevicius%
\thanks{%
Departamento de Matem\'atica and LANA,  Universidade Federal de Santa Maria, Brazil.
E-mail: \url{alicek@ufsm.br}}
\and
R.~J.~Cintra%
\thanks{%
Núcleo de Tecnologia, CAA, Universidade Federal de Pernambuco, Brazil;
previously with the
Departamento de Estat\'istica, Universidade Federal de Pernambuco, Brazil.
E-mail: \url{rjdsc@ufpe.br}}
}

\title{%
An Iterative
Wavelet Threshold for Signal Denoising}

\newcommand{\myabstract}{%
This paper introduces an adaptive filtering process based on shrinking wavelet coefficients from the corresponding signal wavelet representation.
The filtering procedure considers a threshold method determined by an iterative algorithm inspired by the control charts application, which is a tool of the statistical process control (SPC).
The proposed method, called \emph{SpcShrink}, is able to discriminate wavelet coefficients that significantly represent the signal of interest. The \emph{SpcShrink} is algorithmically presented and numerically evaluated according to Monte Carlo simulations. Two empirical applications to real biomedical data filtering are also included and discussed. The \emph{SpcShrink} shows superior performance when compared with competing algorithms.
}

\newcommand{\mykeywords}{%
Adaptive filtering, Control chart, Signal denoising, Wavelet threshold}

\date{}

\begin{document}

\printtitle

\section{Introduction}

Signal denosing is a classical problem
in which a signal is measured in presence of
additive
white
noise~\cite{Nason2011, Elad2006}.
In the past decades numerous contributions
based on different approaches
have been proposed for this problem.
In~\cite{Elad2006},
a list of methods is reviewed:
signal denoising via learned dictionaries~\cite{Elad2006, Aharon2006},
statistical estimators~\cite{Hamza2001, Montanari2015},
spatial adaptive filters~\cite{Woehrle2015,Sanchez2014},
stochastic analysis~\cite{Fouladi2013},
partial differential equations~\cite{Niang2012},
splines and other approximation theory methods~\cite{Unser1999},
morphological analysis~\cite{Singh2006, Maragos1996},
and
transform-domain methods~\cite{Vetterli1992, Donoho1994b, Starck2002, Oppenheim2009, Nason2011}.
In this paper,
we delimit the scope of our contribution to
transform-domain filtering methods
based on discrete wavelet transform (DWT)~\cite{Daub1992, Mallat2008, Nason2011}.

The DWT
is widely regarded as a key tool in
multi-resolution signal analysis~\cite{Daub1992, Mallat2008},
signal detection~\cite{Wang1995, Bailey1998,
Xue2014},
edge detection on images~\cite{Mallat1992,
Zhang2009},
image compression~\cite{Chang2000},
and
signal denoising~\cite{Donoho1994universal, Donoho1995sure, Chang2000, Poorna2008, Oh2009, Srivastava2016},
to mention a few applications in this field.
Among its many possible applications,
denoising techniques have been established as
a major area of signal analysis~\cite{Singer2009},
since
data
is often corrupted by noise during its acquisition or transmission.

Wavelet denoising techniques have been a staple of statistical functional estimation for years  \cite{McGinnity2017}.
These techniques
stem from the pioneer works
by
Donoho and collaborators~\cite{Donoho1994universal, Donoho1995sure},
where theoretical aspects concerning
the application of wavelet transforms
were introduced in this context.
Some recent applications of wavelet denoising
include:
noise removal of biomedical signals~\cite{Kozake2005},
image denoising and compression~\cite{Chang2000},
denoising of hyperspectral imagery~\cite{Chen2011},
despeckling synthetic aperture radar image~\cite{Parrilli2012},
smoothing of quantitative genomic data~\cite{Hatsuda2012},
anomaly detection for mine hunting~\cite{Nelson2012},
seismic noise attenuation~\cite{Goudarzi2012},
and
noise reduction methods for chaotic signals~\cite{Han2013}.
In general,
the central step in wavelet denoising techniques is the
wavelet coefficient shrinkage~\cite{Donoho1995sure}.
This step consists of thresholding or shrinking
the wavelet coefficients in the transform domain.
In signal denoising via wavelet shrinkage,
the threshold value selection is a critical step~\cite{Chen2005}
with
several methods
for
guiding the choice of the threshold value~\cite{Poorna2008}.

In this paper,
we aim at proposing a new denoising scheme
in the class of wavelet-transform methods.
Indeed, we introduce
an innovative approach
for obtaining wavelet threshold values.
The motivation for the proposed threshold value determination
stems from
control chart applications
\cite{Sanusi2017, Hossain2017, Arshad2017},
which
employ statistical process control (SPC)
as
a central tool~\cite{Montgomery2009}.
Control charts are useful
for determining whether a stochastic process
is in the state of statistical control;
being able to distinguish
common and special
variability causes~\cite{Montgomery2009}.
Analogously,
in  the context of signal denoising,
we are interested in identifying
noise (common variability)
and
signal (special variability).
Therefore,
we refer to the sought method as \emph{SpcShrink}.
We introduce a sequence of control limits
that allows
iterative discarding of
wavelet coefficients
until all of them are within a specified control range.
The threshold value
keeps all coefficients inside the control range
after successive limit estimations.
In the \emph{SpcShrink} formulation,
instead of assuming constant variance
for all wavelets coefficients as in~\cite{Donoho1995},
their exponential decay in different scales~\cite{Chen2005}
is considered and
explored in order to compute the desired threshold value.

The proposed method
is presented
and
a
numerical comparison with competing
thresholding schemes in the literature
is provided~\cite{Donoho1994universal,Donoho1995sure,Chang2000,Poorna2008}.
Simulation results
demonstrate the competitive performance of the proposed method.
A computational quality measure assessment
indicates the superiority of the proposed thresholding strategy
when compared with popular methods.
To emphasize the advantages of the proposed method,
real biomedical noisy signals
were filtered according to the \emph{SpcShrink}.
The obtained denoised signal has displayed significant feature preservation.
At the same time,
the \emph{SpcShrink} is capable
of efficiently discarding noise-related information.

This paper unfolds as follows.
Section~\ref{s:algorithm} summarizes
the main concepts of wavelet shrinkage
and SPC;
then
the proposed method is described.
In Section~\ref{ss:choosing},
an optimization problem is solved to specify the optimal control limit distances
for the proposed method.
Numerical experiments are performed in Section~\ref{s:numerical},
comparing the introduced method
with
classical and state-of-the-art wavelet thresholding schemes.
For numerical evaluation
we considered
two real biomedical data
and
Monte Carlo simulations based on synthetic signals.
Conclusions and final remarks
are presented in Section~\ref{s:conclusion}.

\section{SPC-based shrinkage}
\label{s:algorithm}

In this section,
we present a short review on wavelet-based filtering.
Then a summary of the main ideas behind statistical control charts and
hypothesis test is also provided.
Finally,
by combining both tools in a straightforward manner,
we introduce
the \emph{SpcShrink} method.

\subsection{Review of wavelet shrinkage}
\label{s:wave-thresh}

Consider the problem of estimating
an $N$-point unknown signal
$\mathbf{x} = \begin{bmatrix}x[0] & x[1] & \cdots & x[N-1]\end{bmatrix}^\top$
from a set of $N$ noisy observations
$\mathbf{y} = \begin{bmatrix}y[0] & y[1] & \cdots & y[N-1]\end{bmatrix}^\top$
furnished by:
\begin{align}
\label{equation-signal-is-data-plus-noise}
\mathbf{y} = \mathbf{x} + \mathbf{n}
,
\end{align}
where
$\mathbf{n} = \begin{bmatrix}n[0] & n[1] & \cdots & n[N-1]\end{bmatrix}^\top$
is a
white Gaussian noise (WGN) vector
with zero mean and variance $\sigma^2$
($\mathcal{N}(0,\sigma^2)$).
In multiresolution wavelet analysis,
we have $N=2^J$,
where $J$ is the
maximum
number of wavelet decomposition levels~\cite{Daub1992, Mallat2008}.
Let $\mathbf{W}$ be the orthonormal transformation matrix
associated to a given multiresolution wavelet decomposition.
Thus,
the wavelet representation of $\mathbf{y}$ is
given by:
\begin{align*}
\mathbf{w} = \mathbf{W} \cdot \mathbf{y}
.
\end{align*}
In a similar fashion,
we denote
$\mathbf{c} = \mathbf{W} \cdot \mathbf{x}$
and
$\mathbf{z} = \mathbf{W} \cdot \mathbf{n}$.
Because~$\mathbf{W}$
is a linear transformation,
the above quantities
satisfy
$\mathbf{w} = \mathbf{c} + \mathbf{z}$
(cf~\eqref{equation-signal-is-data-plus-noise}).
Additionally,
the orthogonality of the discrete wavelet transform matrix
ensures that
$\mathbf{W}$ transforms white noise into white noise~\cite{Donoho1995sure}.
Thus,
the transformed vector $\mathbf{z}$ is also WGN $\mathcal{N}(0,\sigma^2)$.

For notational purposes,
the entries of vectors
$\mathbf{w}$,
$\mathbf{c}$,
and
$\mathbf{z}$
are doubly indexed
and
denoted as
$w_{j,k}$,
$c_{j,k}$,
and
$z_{j,k}$,
respectively,
where
$j=1,2,\ldots,J$
indicates the scaling domain (associated to frequency)  index
and
$k=1,2,\ldots,2^{J-j}$
denotes the time domain index.

Wavelet shrinkage consists of a judicious
thresholding operation over the elements of~$\mathbf{w}$.
Such operation results in
a modified signal given by
$\hat{\mathbf{w}} = T(\mathbf{w}, \lambda)$,
where
$T(\cdot)$ is the thresholding function
and $\lambda>0$ is the threshold value.
Elements of~$\mathbf{w}$
smaller than $\lambda$
are eliminated or smoothed~\cite{Donoho1995,Chang2000,Smith2008}.
Hard and soft thresholding are common
strategies to `shrink'
wavelet coefficients~\cite{Donoho1995,Chang2000,Smith2008}
and the resulting wavelet coefficients~$\hat{w}_{j,k}$
are,
respectively,
given by:
\begin{align*}
\hat{w}_{j,k}
=&
\begin{cases}
w_{j,k}, & \text{if $|w_{j,k}| > \lambda$,}
\\
0, & \text{otherwise,}
\end{cases}
\\
\hat{w}_{j,k}
=&
\begin{cases}
\operatorname{sign}(w_{j,k}) \cdot (w_{j,k} - \lambda), &
\text{if $|w_{j,k}| > \lambda$,}
\\
0, &
\text{otherwise,}
\end{cases}
\end{align*}
where $\operatorname{sign}(\cdot)$ is the signum function.
Finally,
the true signal
can be estimated based on the shrunk coefficients
according to
$\hat{\mathbf{x}} = \mathbf{W}^\top \cdot \hat{\mathbf{w}}$~\cite{Nason1996}.

Threshold value~$\lambda$ plays a central role in
wavelet shrinkage denoising.
Several
methods for threshold estimation
are described in literature,
such as:
the \emph{VisuShrink} (or universal threshold)~\cite{Donoho1994universal},
the \emph{SureShrink}~\cite{Donoho1995sure},
the \emph{BayesShrink}~\cite{Chang2000},
and
the \emph{S-median}~\cite{Poorna2008}.
Our goal is to propose
an adaptive,
level-dependent method for estimating~$\lambda$
capable of
high performance denoising
and
robust enough for hard or soft thresholding.

\subsection{Review of control charts}\label{s:control-chart}

Control charts are important statistical tools for monitoring disturbances in a statistical process,
and they are richly applied in the industrial sector,
the health sector and the agricultural sector, among others~\cite{Sanusi2017}.
The main objective of control chart application
is the identification
of sources of variability
in manufacturing processes
\cite{Montgomery2009, Oakland2007, Tague2005, Kubiak2009}.
We propose the use of control charts theory
as a means for
threshold value estimation.
A typical control chart
consists of a time series of
statistic measurements
related to the quality characteristic of
a given process.
The chart contains
a center line (CL) representing
the mean value of the considered statistic
and
two horizontal lines,
referred to as
the lower control limit (LCL)
and upper control limit (UCL).

Suppose
normally distributed
quality measurements
with mean $\mu_0$ and standard deviation $\sigma$.
If
$\mathbf{y} = \begin{bmatrix}y[0] & y[1] & \cdots & y[N-1]\end{bmatrix}^\top$
is a vector with $N$ observations of this process,
then
the probability is $1-\alpha$ that any sample
$y[i]$,
with $i=0,1,\ldots,N-1$,
falls within
the following limits:
\begin{align*}
& \mu_0 - d \, \sigma,  \\
& \mu_0 + d \, \sigma
,
\end{align*}
where $d$ is the ($1-\alpha/2$)-quantile of
the standard normal distribution,
or simply $d = \sqrt{2} \operatorname{erfc}^{-1}(\alpha)$,
and
$\operatorname{erfc}^{-1}(\cdot)$
is
the inverse complementary error function~\cite{Carlitz1963,Andrews1992special}.
This way,
the control limits are related according to~\cite{Montgomery2009}:
\begin{align}
\label{equation-limits1}
\mathrm{LCL} =& \mu_0 - d \, \sigma,
\\
\label{equation-limits2}
\mathrm{CL}  =& \mu_0
,
\\
\label{equation-limits3}
\mathrm{UCL} =& \mu_0 + d \, \sigma
.
\end{align}
The quantity $d$
can also be interpreted
as the ``distance'' of the control limits
from its center line,
expressed in standard deviation units.
In control chart applications,
$d$ is usually considered equal to three (three-sigma limits),
where the probability of a simple point
falling outside the limits is $0.0027$~\cite{Montgomery2009, Oakland2007, Tague2005, Kubiak2009, Cook1997}.
In practice,
quantities $\mu_0$ and $\sigma$ are statistically estimated.

There is a close connection between control charts
and hypothesis testing~\cite{Montgomery2009, Kubiak2009}.
If the current value of $y[i]$ is within the control limits,
then the process is considered to be `in control';
that is,
it is an occurrence of a normal distributed variable
with the mean value $\mu_0$.
On the other hand,
if $y[i]$ falls out the control limits,
then
we conclude that the process is `out of control';
that is,
it is an occurrence from a random variable
with a different mean value $\mu_1 \neq \mu_0$.
Therefore
choosing the control limits
is equivalent
to setting up the critical region
for testing the hypothesis~\cite{Montgomery2009}:
\begin{align*}
\begin{split}
\mathcal{H}_0: & \,  \mu=\mu_0 \quad \text{(in control)}
,
\\
\mathcal{H}_1: & \,  \mu \neq \mu_0 \quad  \text{(out of control)}
.
\end{split}
\end{align*}
The control chart tests this hypothesis repeatedly for each observation $y[i]$ of the observed process.
The general procedure in hypothesis testing
starts with the specification
of type~I error $\alpha$ (probability of false alarm),
and then
the design of
a test procedure
that maximizes the power of the test
(probability of detection)~\cite{Montgomery2009}.
Mathematically,
we have
$
\alpha=\operatorname{Pr}(\text{reject $\mathcal{H}_0 | \mathcal{H}_0$ is true})
$
and
$
\text{Power}
=
\operatorname{Pr}
(
\text{reject $\mathcal{H}_0 | \mathcal{H}_0$ is false}
)
$.
Then, by selecting $\alpha$,
we directly control the probability of false alarm.

Practical SPC usage
involves two phases~\cite{Jones2002}:
(i)~identification whether the system is \emph{in control}
and
(ii)~computation of the control limits
to allow in-control
identification.
A system is regarded in control
if most of its related measurements
are
within the control limits~\cite{Oakland2007}.
If a prescribed number of measurements
is outside the range implied by LCL and UCL,
then
such outlying values
are submitted to
investigation
or
exclusion from the analysis set.
If data are discarded,
quantities CL, LCL, and UCL
are recomputed
and
the cycle
is
repeated until all measurements are
within control lines~\cite{Montgomery2009}.
More details about
the statistical analysis of SPC
are found in~\cite{Montgomery2009},
\cite{Oakland2007},
\cite{Tague2005},
and \cite{Kubiak2009}.

\subsection{Proposed \emph{SpcShrink}}\label{s:spc}

Control charts and wavelet shrinkage
share similar goals:
identifying and removing corrupted data---either out of control or
noisy measurements.
We aim at assessing
each wavelet coefficient at a given decomposition level~$j$
to establish whether
it is representative of~$\mathbf{x}$
or
it is simply WGN $\mathbf{n}$.
For such,
we set up the following hypothesis test:
\begin{align}
\label{equation-hypothesis-test}
\begin{split}
\mathcal{H}_0: & \,  c_{j,k}=0
,
\\
\mathcal{H}_1: & \,  c_{j,k}\neq0
.
\end{split}
\end{align}
Notice that
the null hypothesis is linked
to
noise only,
whereas
the alternative hypothesis
indicates the presence of signal embedded in noise.
Therefore,
under the null hypothesis,
we have
$w_{j,k}= z_{j,k}$,
which is WGN $\mathcal{N}(0,\sigma^2)$~\cite{Donoho1995sure},
as discussed in Section~\ref{s:algorithm}.
The normality of the transformed noised data is an important property.
In fact,
it ensures the necessary conditions
for the application
of control chart theory
over the wavelet domain,
as discussed in Section~\ref{s:control-chart}.
A similar
statistical hypothesis formulation
for the wavelet thresholding problem
was
considered in~\cite{Abramovich1996}.

Based on the SPC approach,
we
assess
the probability of
a given wavelet coefficient~$w_{j,k}$
being within the
lower and upper limits based
on~\eqref{equation-limits1}, \eqref{equation-limits2}, and \eqref{equation-limits3}.
Therefore,
for each decomposition level~$j$,
we have that
\begin{align*}
\mathrm{LCL} =& -d_j \cdot s_j,\\
\mathrm{CL} =& 0,\\
\mathrm{UCL} =& d_j \cdot s_j,
\end{align*}
where
$s_j$ is the
corrected
sample standard deviation
of the wavelet coefficients at level~$j$
and
$d_j$ is distance of the control limits
at level~$j$.
In this way,
under the null-hypothesis $\mathcal{H}_0$,
we have that
\begin{align*}
\operatorname{Pr}
\{
| w_{j,k} | \leq d_j \cdot s_j
\}
=
1-\alpha_j
,
\end{align*}
where
$\alpha_j$
are the prescribed
significance levels
and
$d_j = \sqrt{2} \operatorname{erfc}^{-1}(\alpha_j)$.
Indeed,
the quantity
$\alpha_j$
is the probability of detecting a signal when it is not present.
For example,
considering usual values for
the significance level of
$0.2\%$,
$1\%$,
$5\%$,
and
$10\%$
the implied
values for the control limit distance
$d_j$
are
$3.09$,
$2.58$,
$1.96$,
and
$1.64$,
respectively.

\begin{figure*}
\centering
\subfigure[Out of control coefficients ($s=0.93$)]
{\includegraphics[width=0.32\linewidth]{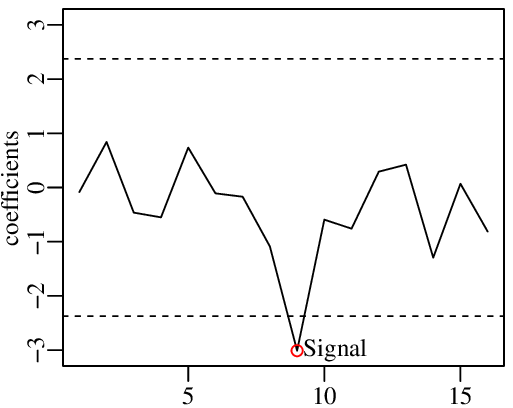}\label{F:phaseI}}
\subfigure[In control coefficients  ($s=0.64$)]
{\includegraphics[width=0.32\linewidth]{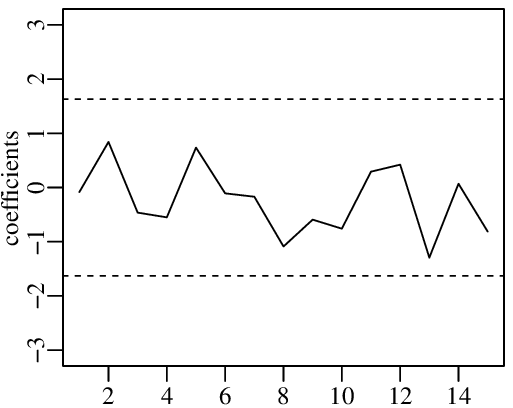}\label{F:phaseII}}
\subfigure[Thresholded coefficients]
{\includegraphics[width=0.32\linewidth]{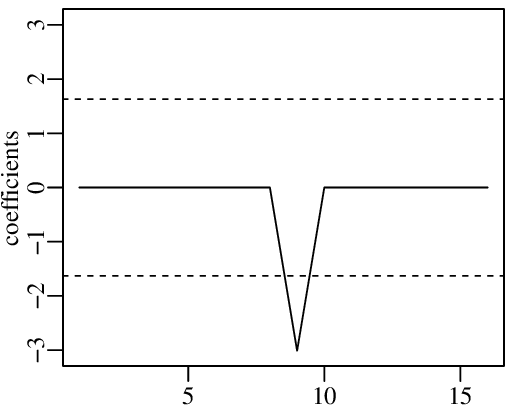}\label{F:phaseIII}}
\caption{Graphical illustration of the proposed iterative method.}
\label{F:method}
\end{figure*}

Thus,
given a limit distance $d_j$,
the proposed
shrinkage method
for estimating threshold values $\lambda_j$
at scale level~$j$
is given by
the iterative procedure described as follows:
\begin{enumerate}
\item
Estimate the standard deviation of the wavelet coefficients at level $j$:
\begin{align*}
s_j
=
\sqrt{
\frac{1}{N_j-1}
\sum_{k=1}^{N_j}
(w_{j,k}-\bar{w}_j\,)^2
}
,
\end{align*}
where
$N_j=2^{J-j}$ is the number of wavelet coefficients at multi-resolution level $j$
and
$\bar{w}_j=\frac{1}{N_j}\sum\limits_{k=1}^{N_j} w_{j,k}$.

\item
Establish the control limits according to:
\begin{align*}
\mathrm{LCL}=& -d_j \cdot s_j,
\\
\mathrm{UCL}=& \; d_j \cdot s_j,
\end{align*}
where
$d_j = \sqrt{2} \operatorname{erfc}^{-1}(\alpha_j)$
for the prescribed significance levels~$\alpha_j$.
Figure~\ref{F:method} illustrates the concept.

\item
If a wavelet coefficient $w_{j,k}$, for $k=1,2,\ldots,N_j$,
exceeds the control limits $[\mathrm{LCL}, \mathrm{UCL}]$,
then this coefficient is excluded
and Steps~1) and~2) are repeated (Figure~\ref{F:phaseII}).
If all $w_{j,k}$ are inside the interval $[\mathrm{LCL}, \mathrm{UCL}]$,
then the threshold value is $\lambda_j =d_j \cdot s_j$
(Figures~\ref{F:phaseII} and \ref{F:phaseIII})
and the iterative method is stopped.

\end{enumerate}

It is a well-known fact
that noise tends
to be more pronounced
at the lower scale levels
and
becomes less pronounced as the scale level $j$ increases~\cite{Chen2005}.
Thus,
as $j$ increases,
the significance level $\alpha_j$
can
also increase in value.
The increase of
the
type~I error probability $\alpha_j$
along the scales $j$
improves
the
probability of detection
 (power of test)
of
the hypothesis test
described in~\eqref{equation-hypothesis-test}.
Therefore,
in view of this behavior,
the control limit distance
must be adjusted accordingly.
An increase in $\alpha_j$
effects a decrease in the value
of the corresponding
control limit distance~$d_j$,
because the function $\operatorname{erfc}^{-1}(\cdot)$
is a monotonically decreasing function over
the interval $[0, 1]$.

In a conservative fashion,
we adopt a simple
linear increase of $\alpha_j$ along the scales
according to:
\begin{align*}
\alpha_j = j \cdot \alpha_1,
\quad
j=2,3,\ldots,J_0
,
\end{align*}
where
$J_0 \leq J$ is the number of decomposition levels
and
$\alpha_1$ is
a statistical significance level
to be determined.
As a consequence,
we have that the control limit distances are
given by
$d_j = \sqrt{2} \cdot \operatorname{erfc}^{-1}\left\{j \cdot \alpha_1 \right\}$.
Thus,
because
the proposed shrinkage method depends on the $\alpha_1$ value,
we refer to it as \emph{SpcShrink}($\alpha_1$).
Thus,
our proposed method has a free parameter
in a similar way
as the \emph{S-median} method proposed by \cite{Poorna2008}.
In the Section~\ref{ss:choosing} we propose optimum values for $\alpha_1$.

\subsection{Adaptivity and convergence}

The proposed iterative method is adaptive in two senses:
(i)~it considers different threshold values for each wavelet decomposition level
and
(ii)~based on control chart arguments,
it discards detected signal samples
and recalculates the threshold estimates,
adapting itself to the signal characteristics.

The asymptotic convergence of the method is guaranteed.
As it happens in statistical control charts,
the \emph{SpcShrink}
can be understood
as a heuristic to derive two sequences
$\text{LCL}_i$
and
$\text{UCL}_i$
being
$\text{LCL}_i \leq \text{UCL}_i$,
for each iteration $i$.
For each iteration,
these sequences are based on the computation
of the mean value of the current data set,
which is bounded
by the maximum/minimum values of the data set.
Each iteration discards some samples
and the mean value is recomputed.
This means the \emph{SpcShrink} method,
for each level $i$,
generates two limited and monotonic
sequences of real values;
thus they must converge to a value
when the
number of iterations goes to infinity~\cite{Laczkovich2015}.

Let $J_0 \leq J$
be the selected number of decomposition levels.
For the
algorithm introduced in Section~2.3,
the numerical convergence is guaranteed.
The main mathematical operations specifically required
by the proposed algorithm
are:
(i)~standard deviation computation (Step 1);
(ii)~evaluation of upper and lower control limits (Step 2);
and
(iii)~element removal from a given array (Step 3).
These operations require, respectively:
$J_0$ calls per iteration;
$J_0$ multiplications per iteration;
and
$J_0$ calls per iteration of find/sort algorithms
which are computed in $\mathcal{O}(N \cdot \log N)$
by quicksort variants~\cite{Cormen2009}.
Indeed,
in the worst case scenario,
each iteration
discards
only
a single wavelet coefficient.
Since each decomposition level
must keep at least one coefficient,
the number of iterations
is upper bound by $N-J_0$.
Common to shrinkage algorithms,
we have two calls (forward and inverse)
of the particular wavelet transform at hand
whose complexity is in $\mathcal{O}(N)$.
Therefore,
the proposed algorithm
can be efficiently implemented in contemporary software packages
at a very low computational cost.

\section{Optimum specification of the control limit distances}
\label{ss:choosing}

The proposed method,
as any hypothesis test,
depends on the significance levels
that define
the control limit distances ($d_j$).
In this section,
we propose and solve an optimization problem to identify suitable
values of a free parameter $\alpha_1$ that defines the limits
$d_j = \sqrt{2} \cdot \operatorname{erfc}^{-1}\left\{j \cdot \alpha_1 \right\}$.
Computational and qualitative analyses are also presented.

\subsection{Optimization problem}

To obtain
the optimal value for the significance levels,
we introduce the following optimization problem:
\begin{align}
\label{equation-optimization}
\alpha_1^\ast
=
\arg
\min_{\alpha_1 \in A}
\frac{1}{M}
\sum_{i=1}^M
\operatorname{Error}
(\mathbf{x}_i, \hat{\mathbf{x}}_i)
,
\end{align}
where
$\mathbf{x}_i$ is an input signal,
$\hat{\mathbf{x}}_i$ is the associated denoised signal
according to the proposed scheme,
$\operatorname{Error}(\cdot,\cdot)$
is a figure of merit to assess the denoised signal,
$M$ is the number of signal instantiations,
and
$\mathcal{A}$ is the search space.
For
the
error measure $\operatorname{Error}$,
we adopted
the
negative
value of the classical signal-to-noise ratio (SNR).
This measure is defined by
\begin{align}\label{e:snr}
\operatorname{SNR}
(
\mathbf{x}_i, \hat{\mathbf{x}}_i
)
=
10
\log_{10}
\left(
\frac{\sigma^2_\text{signal}}{\sigma^2_\text{noise}}
\right)
,
\end{align}
where
$\sigma^2_{\text{signal}}$ is the variance of the $\mathbf{x}_i$
and
$\sigma^2_{\text{noise}}$ is the variance of $\mathbf{x}_i - \hat{\mathbf{x}}_i$.
In practice,
variance estimates are considered.
The SNR is given in decibels (dB),
where greater SNR value indicate
better filtering.

\subsection{Computational search}

\begin{figure*}%
\centering
\subfigure[Blocks]
{\includegraphics[width=0.32\linewidth]{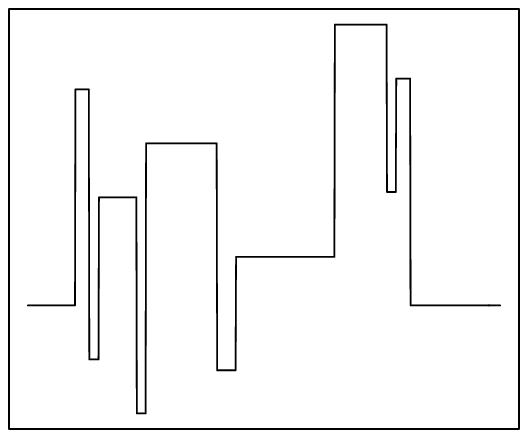} \label{F:blocks}}
\subfigure[Bumps]
{\includegraphics[width=0.32\linewidth]{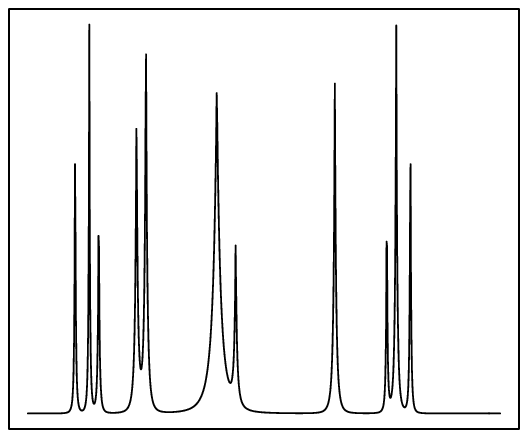} \label{F:bumps}}
\subfigure[Doppler]
{\includegraphics[width=0.32\linewidth]{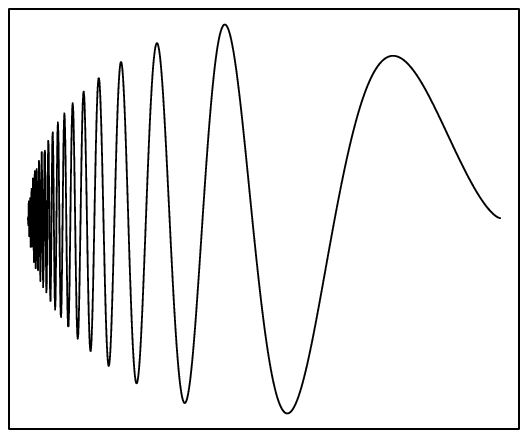} \label{F:doppler}}
\caption{Standardized signals used in the proposed numerical evaluations.}
\label{F:signals}
\end{figure*}

\begin{figure}%
\centering
{\includegraphics[width=0.5\linewidth]{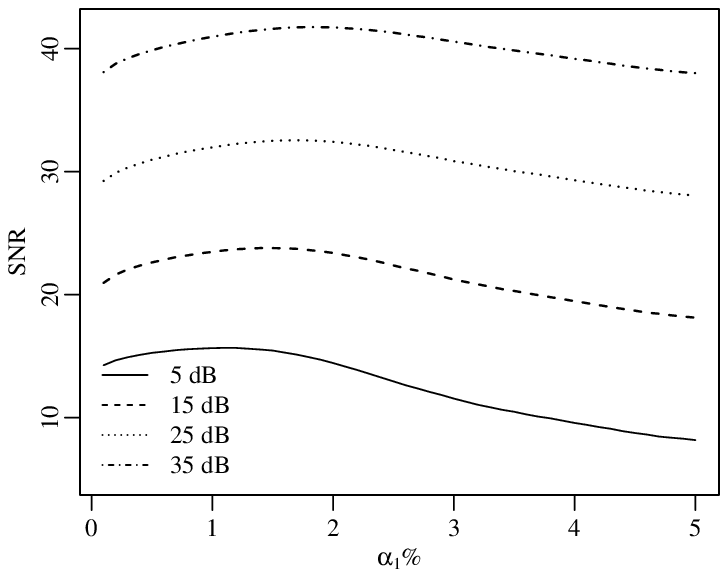} }
\caption{Quantitative \emph{SpcShrink}($\alpha_1$) comparison considering several $\alpha_1$ values for different levels of corrupted signal (in dB).
 }
\label{figure-p1}
\end{figure}

In order to solve~\eqref{equation-optimization},
we set up
a Monte Carlo simulation~\cite{Doucet2005, Rizzo2007}.
For input data,
we separated the signals depicted in Figure~\ref{F:signals}.
These are standard signals largely employed
for assessing filtering processes
as shown in~\cite{Donoho1995sure,Donoho1994universal}
and \cite{Hurvich1998}.
The selected signal blocklength was $2^{12}$ points.
As suggested in~\cite{Chang2000} and \cite{Poorna2008},
we adopted the Daubechies wavelet
with eight vanishing moments~\cite{Daub1992}
as the analyzing wavelet.
Five scales
($J_0=5$)
were considered in the orthogonal wavelet decomposition and
soft thresholding was employed.
Considering $M=1500$ replications of the standard signals,
500~instantiations for each signal shown
in Figure~\ref{F:signals},
we submitted them
to
additive WGN
with different variances $\sigma^2_{\text{noise}}$
resulting
in the following SNR:
5, 15, 25, and 35~dB,
according to~\eqref{e:snr}.

We adopted
the following search space:
$\mathcal{A} = \{0.1\%, 0.2\%, 0.3\%, \ldots, 4.9\%, 5\%\}$.
The resulting denoised signals
were assessed by means of
SNR
measurements.
Results were averaged and
the numerical minimum was sought
(cf.~\eqref{equation-optimization}).
Figure~\ref{figure-p1}
displays SNR
plots
over~$\mathcal{A}$.
For each considered
noise level,
a different value of $\alpha_1^\ast$
was obtained.
The obtained minima
are shown in Table~\ref{table-optimal-values}.
Because the quantity~$\alpha_1^\ast$ varies
according to the injected noise level,
we
computed
the average of the minima.
Hereafter,
we refer to such average value simply as $\alpha_1$.
Thus,
we identified the mean value $\alpha_1 = 1.5\%$
that represent a good compromise solution
for all considered scenarios.
For $\alpha_1=1.5\%$ the control limit distances
are
$d_1= 2.432$,
$d_2= 2.170$,
$d_3= 2.005$,
$d_4= 1.881$,
and
$d_5= 1.780$.

\begin{table}[h]
\centering
\caption{Optimal parameters $\alpha_1^\ast$}
\label{table-optimal-values}
\begin{tabular}{c|cccc|c}
\toprule
\cmidrule(r){2-6}
Noise level  & $5$ dB & $15$ dB & $25$ dB & $35$ dB & Mean
\\
\midrule
$\alpha_1^\ast$
&
1.1 & 1.4 & 1.7 & 1.8 & 1.5
\\
\bottomrule
\end{tabular}
\end{table}

\subsection{Qualitative analysis}

\begin{figure*}%
\centering
\vspace{-0.2cm}
\subfigure[Input blocks signal with 20 dB]
{\includegraphics[width=0.25\linewidth]{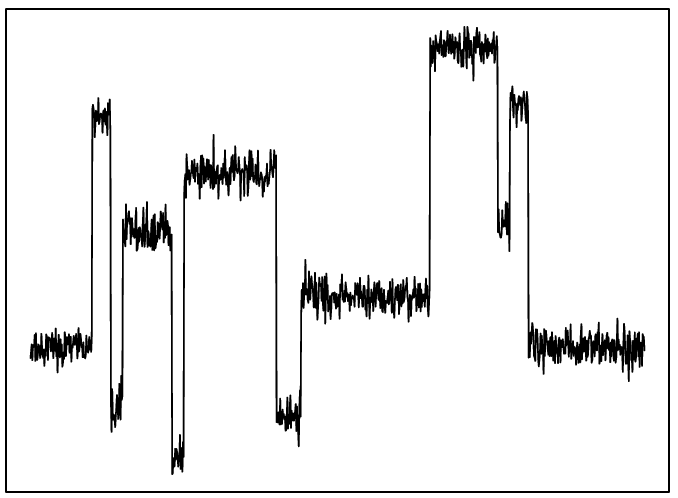}}
\subfigure[Input bumps signal with 20 dB]
{\includegraphics[width=0.25\linewidth]{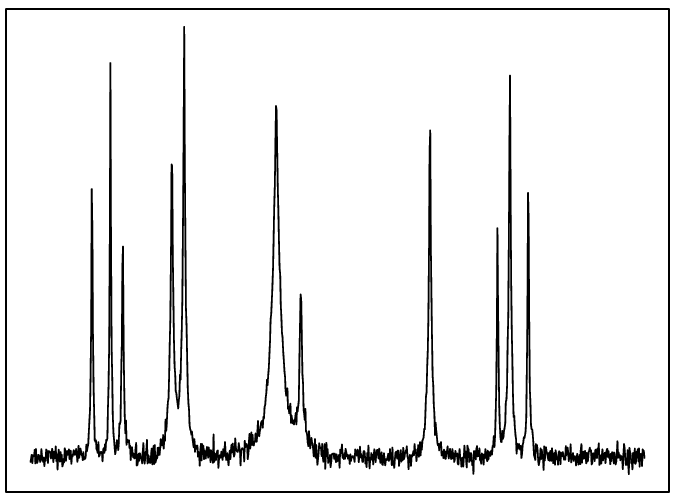}}
\subfigure[Input Doppler signal with 20 dB]
{\includegraphics[width=0.25\linewidth]{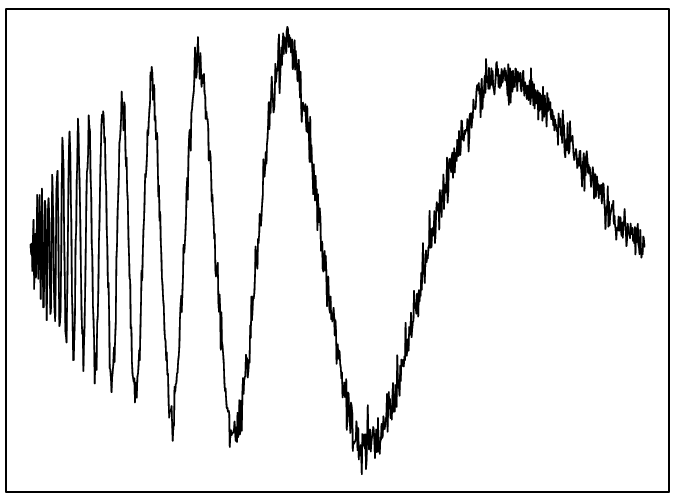}}
\subfigure[$\alpha_1\!=\!5\%$ (SNR=24.17)]
{\includegraphics[width=0.25\linewidth]{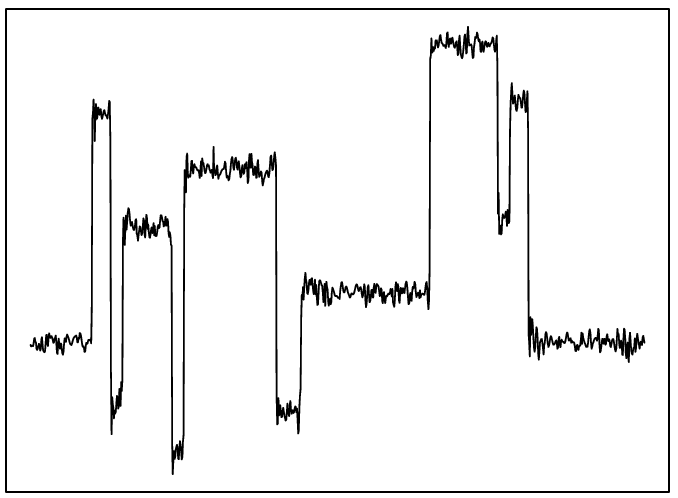}}
\subfigure[$\alpha_1\!=\!5\%$ (SNR=24.13)]
{\includegraphics[width=0.25\linewidth]{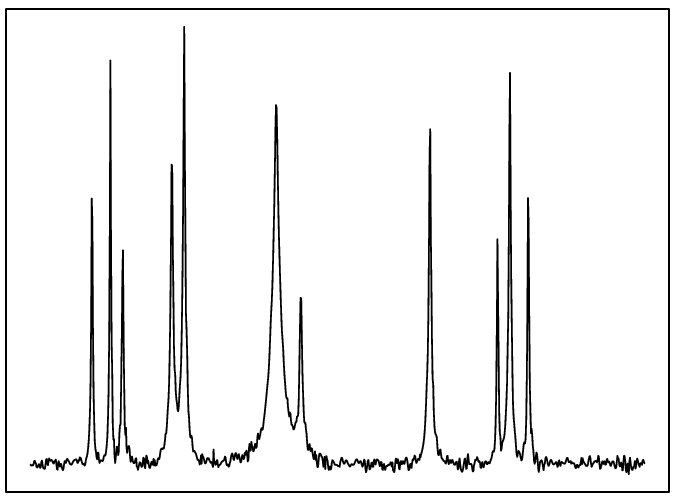}}
\subfigure[$\alpha_1\!=\!5\%$ (SNR=22.92)]
{\includegraphics[width=0.25\linewidth]{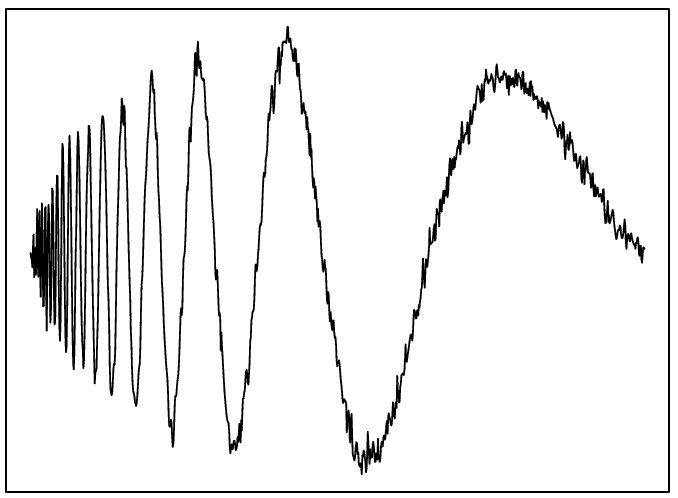}}
\subfigure[$\alpha_1\!=\!2.5\%$ (SNR=26.24)]
{\includegraphics[width=0.25\linewidth]{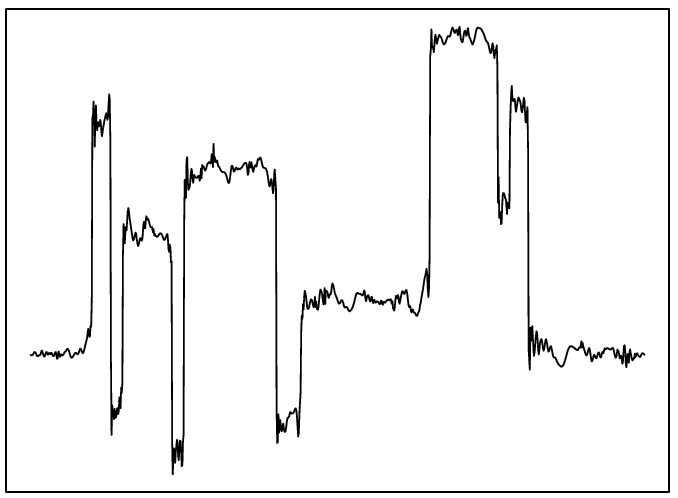}}
\subfigure[$\alpha_1\!=\!2.5\%$ (SNR=26.97)]
{\includegraphics[width=0.25\linewidth]{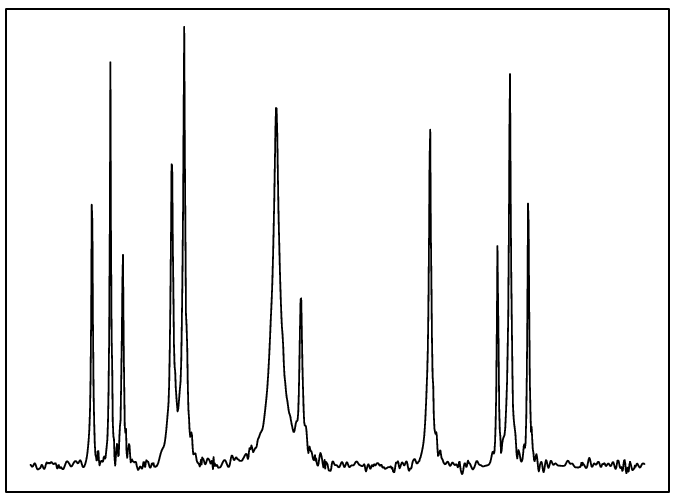}}
\subfigure[$\alpha_1\!=\!2.5\%$ (SNR=25.65)]
{\includegraphics[width=0.25\linewidth]{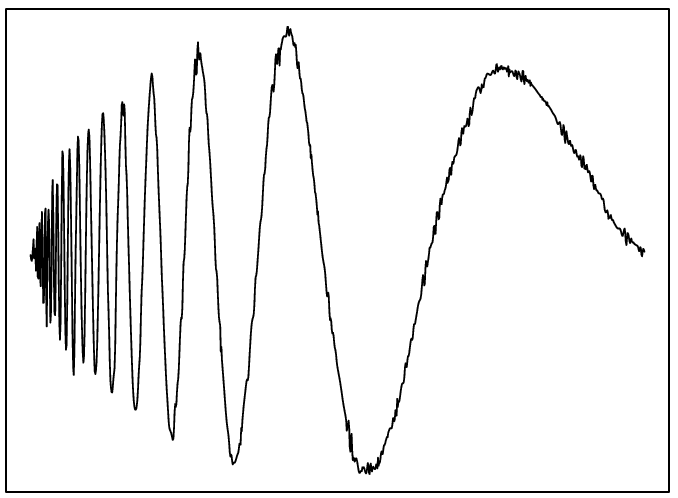}}
\subfigure[$\alpha_1\!=\!1.5\%$ (SNR=26.61)]
{\includegraphics[width=0.25\linewidth]{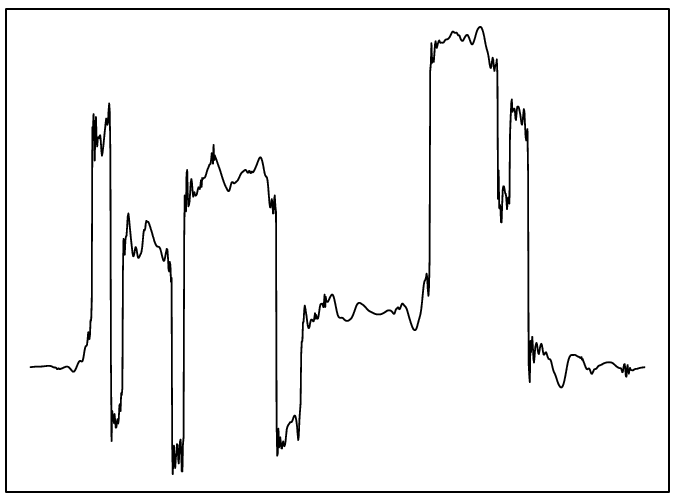}}
\subfigure[$\alpha_1\!=\!1.5\%$ (SNR=27.28)]
{\includegraphics[width=0.25\linewidth]{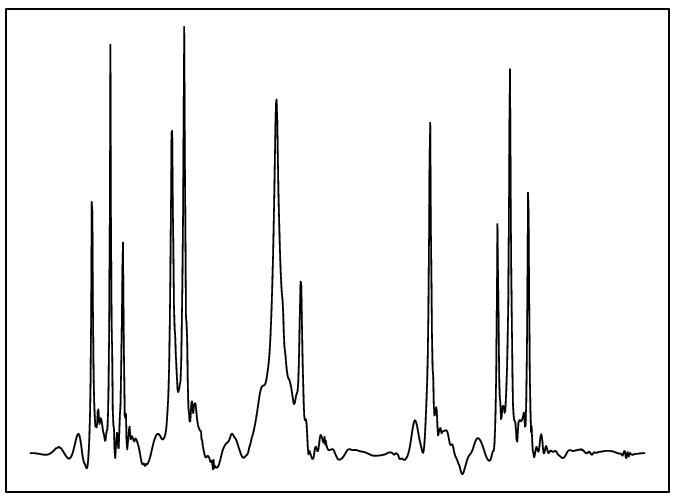}}
\subfigure[$\alpha_1\!=\!1.5\%$ (SNR=25.63)]
{\includegraphics[width=0.25\linewidth]{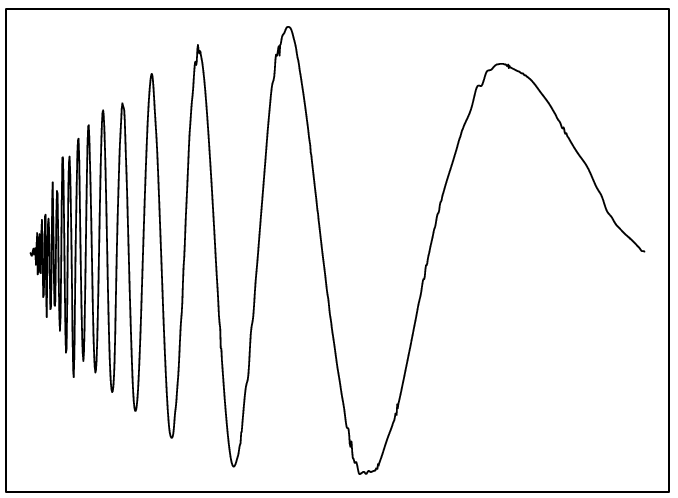}}
\subfigure[$\alpha_1\!\!=\!\!1.0\%$ (SNR=25.57)]
{\includegraphics[width=0.25\linewidth]{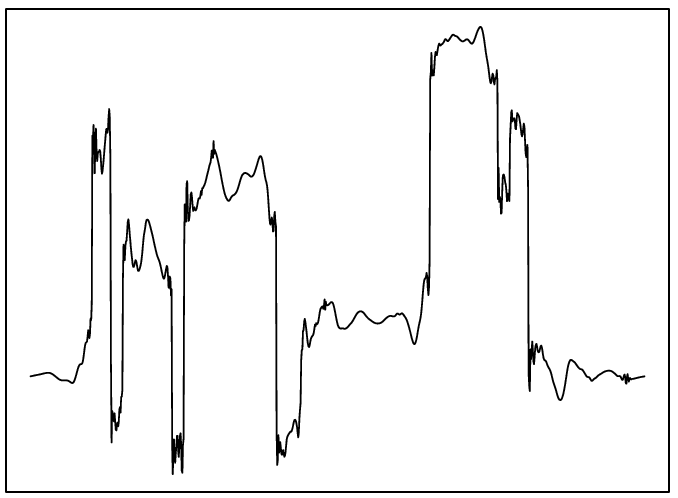}}
\subfigure[$\alpha_1\!=\!1.0\%$ (SNR=26.37)]
{\includegraphics[width=0.25\linewidth]{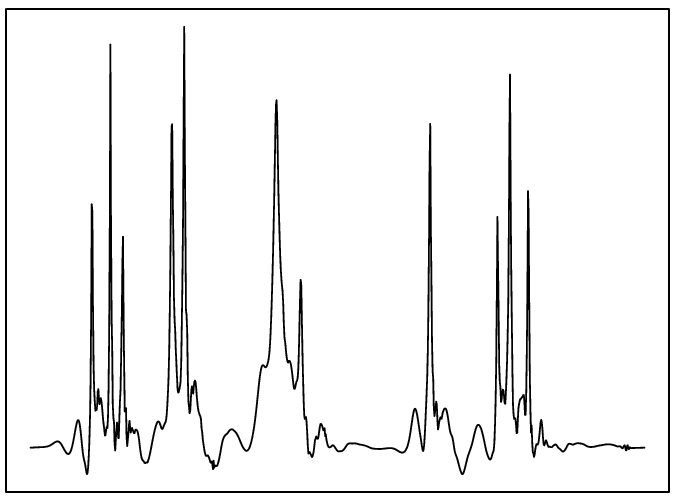}}
\subfigure[$\alpha_1\!=\!1.0\%$ (SNR=25.03)]
{\includegraphics[width=0.25\linewidth]{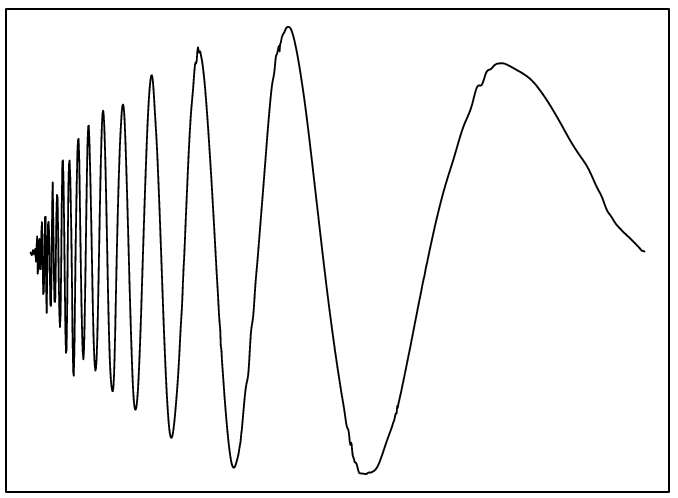}}
\subfigure[$\alpha_1\!=\!0.1\%$ (SNR=22.46)]
{\includegraphics[width=0.25\linewidth]{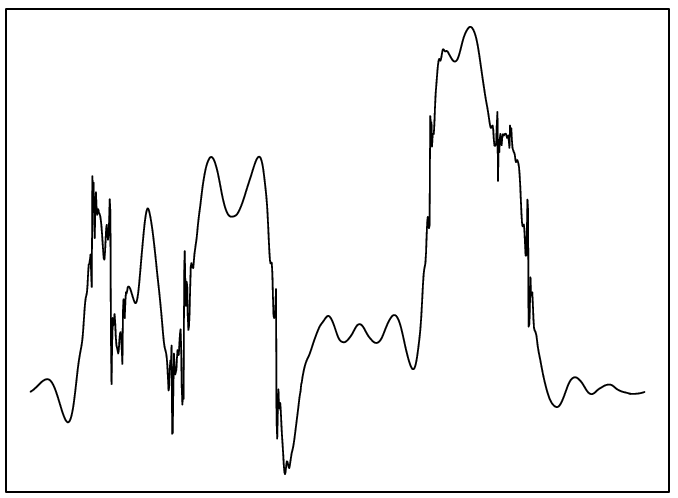}}
\subfigure[$\alpha_1\!=\!0.1\%$ (SNR=22.97)]
{\includegraphics[width=0.25\linewidth]{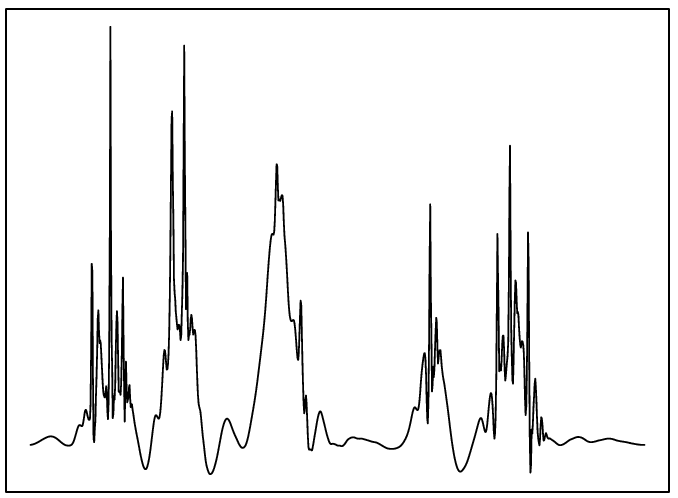}}
\subfigure[$\alpha_1\!=\!0.1\%$ (SNR=22.31)]
{\includegraphics[width=0.25\linewidth]{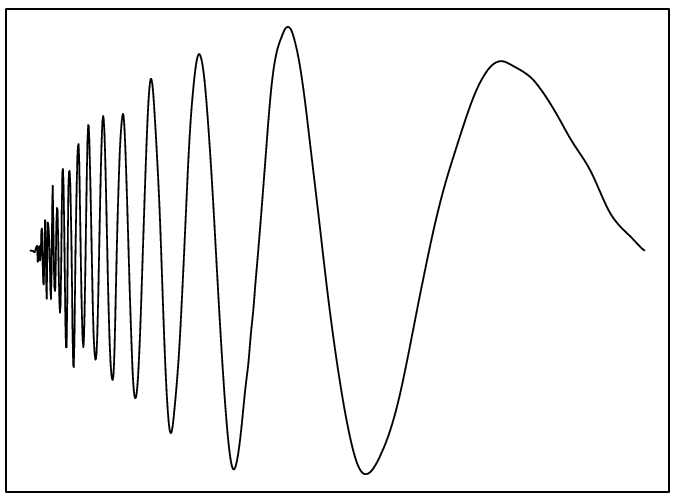}}
\vspace{-0.2cm}
\caption{Output filtered signal comparison by \emph{SpcShrink}($\alpha_1$) with several $\alpha_1$ values.}
\label{F:spc_comparison}
\end{figure*}

Filtering methods that minimize a quantitative measure,
such as SNR or mean square error,
do not necessarily lead in better image visual quality results
in the sense described by
\cite{Donoho1994b},
\cite{Donoho1995},
and \cite{Donoho1994universal}.
As an example,
the \emph{VisuShrink}~\cite{Donoho1994universal}
provides better visual quality
than
procedures based on
mean square error minimization~\cite{Donoho1995}.
Therefore,
we
examined the
behavior of the proposed method
at the vicinity of
the optimal value $\alpha=1.5\%$.
The goal
is to probe
for
resulting signals
which
display
good visual quality
after
wavelet coefficient shrinkage.

Figure~\ref{F:spc_comparison}
shows a qualitative analysis
of the denoised signals
according to the proposed method
for
noisy signals with SNR of 20~dB
and
several values of $\alpha_1$.
Notice that for $\alpha_1 = 1.5\%$,
in general,
the discussed thresholding
yielded
higher SNR output signal.
On the other hand,
the case where  $\alpha_1 = 1.0\%$ provides better visual result.
Similar to \emph{VisuShrink},
\emph{SpcShrink}($1.0\%$)
furnishes outputs with smooth visual appearance
at a lower SNR.
We note that as the value of $\alpha_1$
decreases
the output signal becomes smoother.

The proposed method
is capable of trading-off quantitative measurements (e.g., SNR)
for smoothness (cf.~\emph{VisuShrink}~\cite{Donoho1994universal}).
Such balancing is not available in
traditional
wavelet shrinkage
methods,
such as
\emph{VisuShrink},
\emph{SureShrink},
and
\emph{BayesShrink}.

\section{Numerical comparisons}
\label{s:numerical}

In this section we aim at comparing
the proposed method with the following well-known shrinkage schemes:
\emph{VisuShrink}~\cite{Donoho1994universal},
\emph{SureShrink}~\cite{Donoho1995sure},
\emph{BayesShrink}~\cite{Chang2000},
and
\emph{S-median}~\cite{Poorna2008}.
For the introduced
scheme,
we selected the
control limit distances
associated to
$\alpha_1 \in \{ 1.0\%, 1.5\% \}$,
because these values of $\alpha_1$
were demonstrated to lead to useful results
as described in the previous section.

We present computational experiments
to assess the performance of the
proposed \emph{SpcShrink}($\alpha_1$) filtering process
in two scenarios:
(i)~synthetic signals (Section~\ref{s:monte-carlo})
and
(ii)~actual biomedical data (Section~\ref{s:application}).
For data in Scenario~(i),
we evaluated
the influence of the noise level in denoising techniques
by means of Monte Carlo simulations.
Additionally,
the impact of changing the wavelet filter length
is also stressed in this section.
In Scenario~(ii),
we also include
a visual quality assessment
of the filtered data,
since this is a relevant aspect
in applications involving actual biomedical signals~\cite{aldroubi1996}.
All computations
were
performed
in the \texttt{R}~programming language~\cite{R2013}.

\subsection{Synthetic data denosing}\label{s:monte-carlo}

\begin{figure*}[t]
\centering
\subfigure[Blocks]
{\includegraphics[width=0.32\linewidth]{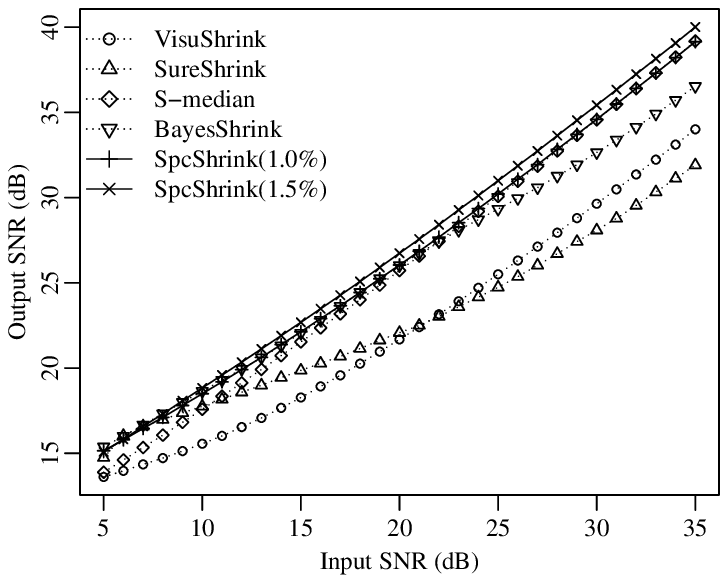} \label{F:blocks_snr}}
\subfigure[Bumps]
{\includegraphics[width=0.32\linewidth]{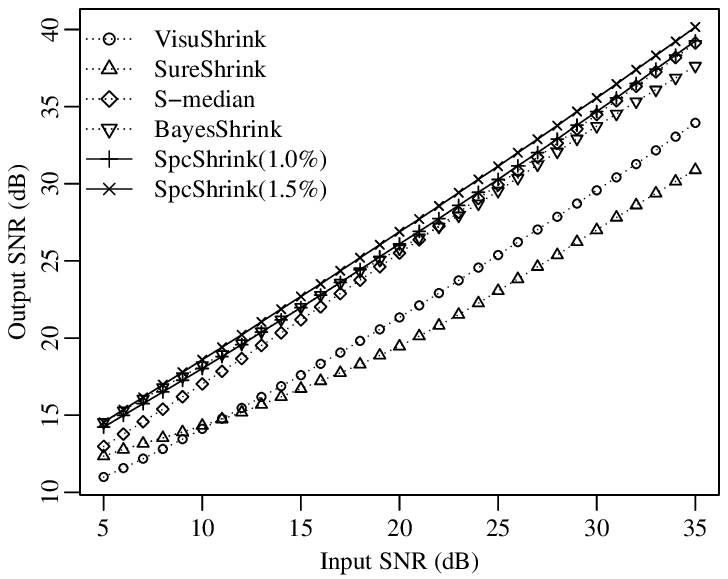} \label{F:bumps_snr}}
\subfigure[Doppler]
{\includegraphics[width=0.32\linewidth]{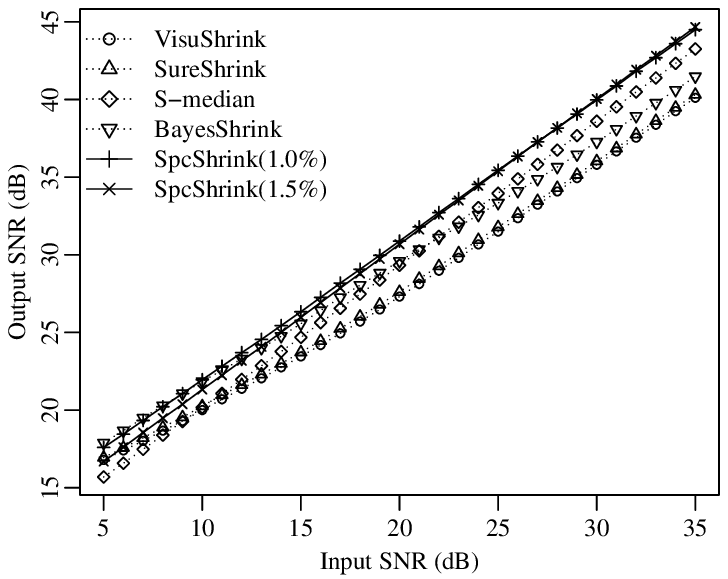} \label{F:doppler_snr}}
\caption{SNR measures comparison of different thresholding for several noise level.}
\label{F:snr}
\end{figure*}

\begin{figure*}[t]
\centering
\subfigure[Blocks]
{\includegraphics[width=0.32\linewidth]{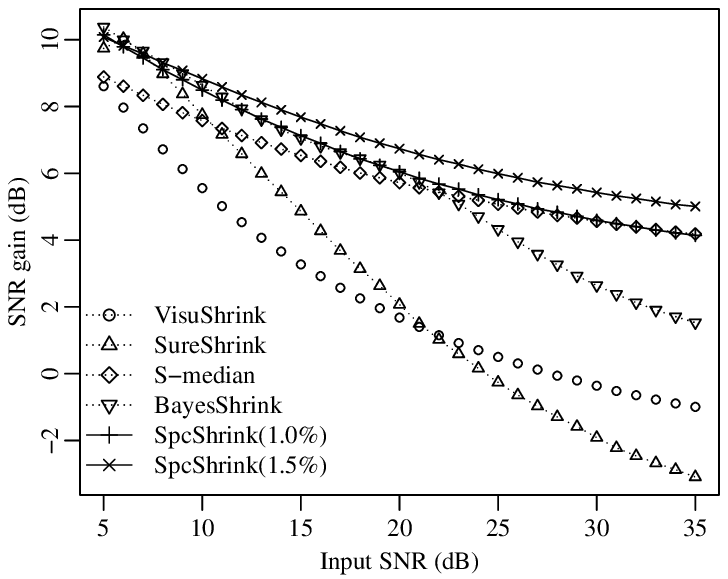} \label{F:blocks_snr_gain}}
\subfigure[Bumps]
{\includegraphics[width=0.32\linewidth]{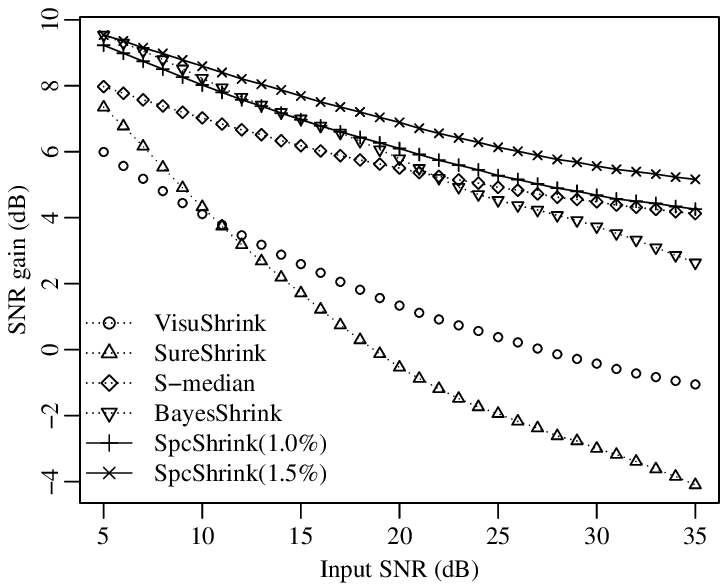} \label{F:bumps_snr_gain}}
\subfigure[Doppler]
{\includegraphics[width=0.32\linewidth]{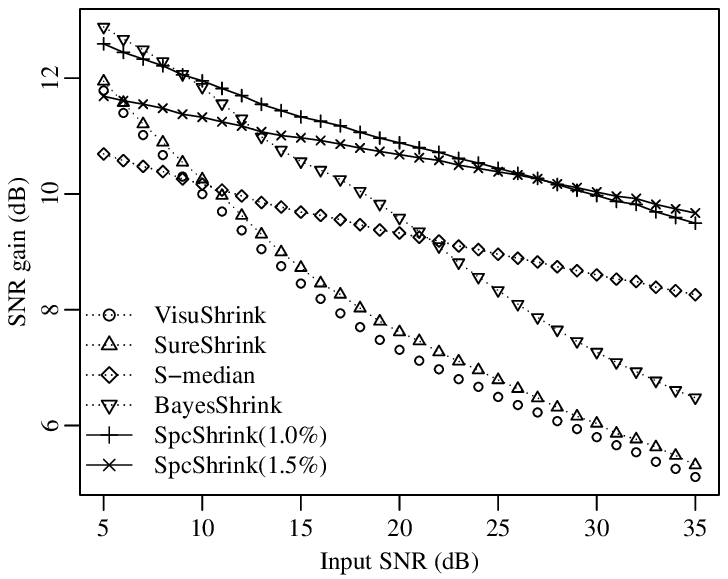} \label{F:doppler_snr_gain}}
\caption{Quantitative noise suppression analysis of different thresholding in SNR gain.}
\label{F:snr_gain}
\end{figure*}

\begin{figure*}[t]
\centering
\subfigure[Blocks]
{\includegraphics[width=0.32\linewidth]{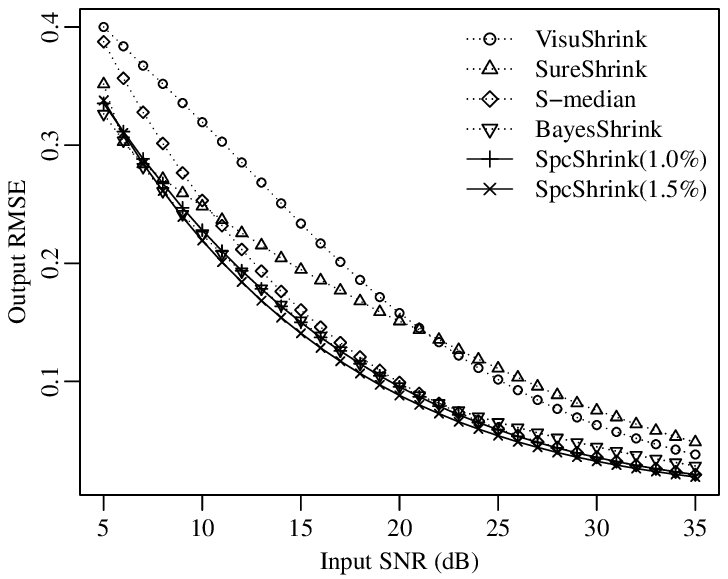} \label{F:blocks_rms}}
\subfigure[Bumps]
{\includegraphics[width=0.32\linewidth]{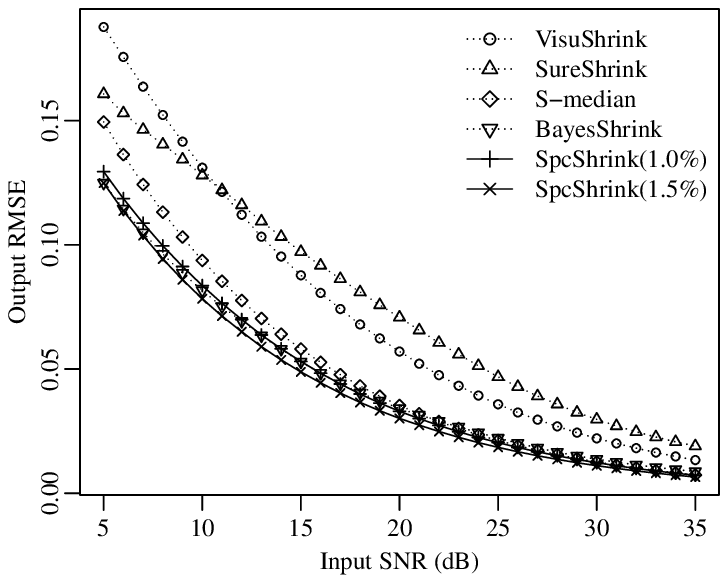} \label{F:bumps_rms}}
\subfigure[Doppler]
{\includegraphics[width=0.32\linewidth]{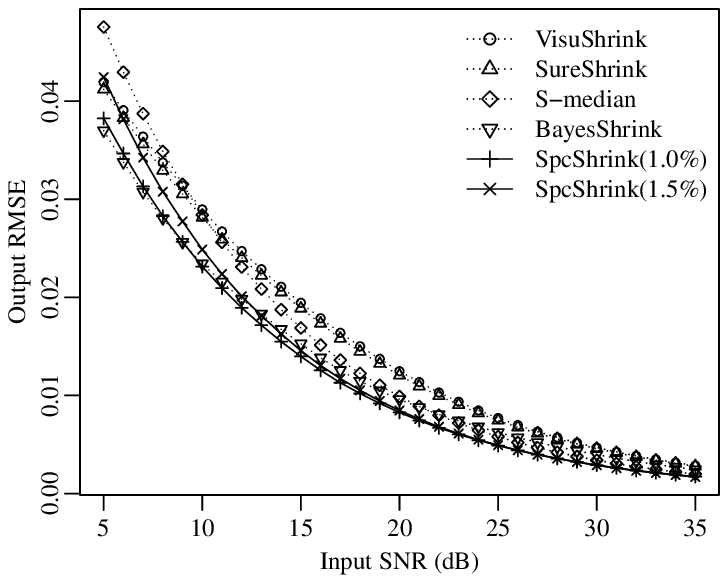} \label{F:doppler_rms}}
\caption{RMSE measures comparison of different thresholding for several noise levels.}
\label{F:rms}
\end{figure*}

Firstly,
to
assess the signal denoising performance,
the SNR
was
employed as figure of merit.
As considered in \cite{Poorna2008},
we also employed the root mean square error (RMSE) for this evaluation.
SNR and RMSE values were computed in average;
considering 1{,}000 replications of the described Monte Carlo simulation.
For this simulation
we adopted the Daubechies wavelet
with eight vanishing moments
and
five scales ($J_0=5$).
The selected blocklength of the signals in Figure~\ref{F:signals} was $2^{12}$ points.
Following an approach similar to the one described in~\cite{Poorna2008},
we fixed the
WGN levels
to signals with input SNR varying
from 5~dB to 35~dB
in steps of
1~dB,
as shown in Figures~\ref{F:snr} and~\ref{F:rms}.
Thus,
we considered
three benchmarking signals,
additive WGN under 31 different variance levels,
and 1,000 Monte Carlo replications,
totalizing
$(3 \times 31 \times 1,000)=93,000$
simulated signals
in our validation experiments.

Considering the SNR results presented in Figure~\ref{F:snr},
the proposed scheme
outperforms
the competing methods
in almost every analyzed scenarios,
except for the `Doppler' input signal with noise level smaller than 10 dB.
For the `blocks' and `bumps' signals
the \emph{SpcShrink}(1.5\%)
furnished best results,
whereas
for the `Doppler' signal
the \emph{SpcShrink}(1.0\%)
is recommended.
As suggested in~\cite{aldroubi1996},
we also considered the SNR gain
of the proposed method.
The SNR gain is
the difference between the output SNR and input SNR values.
Therefore,
the SNR gain
can quantify
the noise suppression.
Figure~\ref{F:snr_gain}
shows the results.
The proposed method
achieves the highest gains
in almost all considered scenarios.
In particular,
gains of more than 12~dB could be
attained
when the input SNR
is lower than 10~dB.

\begin{table*}[ph]
\centering
\caption{SNR comparison of the wavelet denoising methods with different scales ($J_0$) in wavelet decomposition for some corrupted signals
}
\label{t:results}
\begin{tabular}{cccccccc}
\toprule
Signal	&\!\!\!\!	Input SNR	&	\!\!\!\!\emph{VisuShrink}	&	\!\!\!\!\emph{SureShrink}	& \!\!\!\!	\emph{S-median}	&\!\!\!\!	\emph{BayesShrink}	& \!\!\!\!\!	\emph{SpcShrink(1.0\%)}	& \!\!\!\!\!	\emph{SpcShrink(1.5\%)}	\\
\midrule
\multicolumn{8}{c}{$J_0=3$}\\
\midrule
	& $	10	$ & $ 	16.9284	$ & $ 	17.1682	$ & $ 	17.1529	$ & $ 	17.3786	$ & $ 	\mathbf{17.5157}	$ & $ 	17.4955	$ \\
Blocks	& $	20	$ & $ 	23.6527	$ & $ 	23.4104	$ & $ 	25.8875	$ & $ 	25.6691	$ & $ 	25.8675	$ & $ 	\mathbf{26.1224}	$ \\ \vspace{0.15cm}
	& $	30	$ & $ 	31.4303	$ & $ 	30.7546	$ & $ 	34.9138	$ & $ 	32.5868	$ & $ 	34.7118	$ & $ 	\mathbf{35.1352}	$ \\
	& $	10$ & $ 	16.6927	$ & $ 	16.6328	$ & $ 	17.0943	$ & $ 	17.5152	$ & $ 	17.5620	$ & $ 	\mathbf{17.5748}	$ \\
Bumps	& $	20	$ & $ 	23.9173	$ & $ 	23.5005	$ & $ 	25.9814	$ & $ 	25.6219	$ & $ 	26.1922	$ & $ 	\mathbf{26.4381}	$ \\ \vspace{0.15cm}
	& $	30	$ & $ 	31.7294	$ & $ 	31.0382	$ & $ 	35.0473	$ & $ 	33.6996	$ & $ 	34.9675	$ & $ 	\mathbf{35.3802}	$ \\
	& $	10	$ & $ 	18.7457	$ & $ 	18.7465	$ & $ 	18.0588	$ & $ 	\mathbf{18.7533}	$ & $ 	18.6386	$ & $ 	18.4332	$ \\
Doppler	& $	20	$ & $ 	27.7186	$ & $ 	27.7476	$ & $ 	27.7750	$ & $ 	28.3486	$ & $ 	\mathbf{28.3698}	$ & $ 	28.2441	$ \\
	& $	30	$ & $ 	36.9345	$ & $ 	36.9477	$ & $ 	37.4739	$ & $ 	36.9797	$ & $ 	\mathbf{38.0408}	$ & $ 	37.9911	$ \\
	\midrule
\multicolumn{8}{c}{$J_0=5$}\\
\midrule
	& $	10	$ & $ 	15.5543	$ & $ 	17.7555	$ & $ 	17.5772	$ & $ 	18.6515	$ & $ 	18.4976	$ & $ 	\mathbf{18.8388}	$ \\
Blocks	& $	20	$ & $ 	21.6718	$ & $ 	22.0240	$ & $ 	25.7155	$ & $ 	26.0159	$ & $ 	26.0451	$ & $ 	\mathbf{26.7315}	$ \\ \vspace{0.15cm}
	& $	30	$ & $ 	29.6308	$ & $ 	28.0889	$ & $ 	34.5558	$ & $ 	32.6437	$ & $ 	34.6019	$ & $ 	\mathbf{35.4334}	$ \\
	& $	10	$ & $ 	14.1192	$ & $ 	14.2867	$ & $ 	17.0311	$ & $ 	18.2352	$ & $ 	18.0367	$ & $ 	\mathbf{18.5925}	$ \\
Bumps	& $	20 $ & $ 	21.3501	$ & $ 	19.4759	$ & $ 	25.5057	$ & $ 	25.7783	$ & $ 	26.1079	$ & $ 	\mathbf{26.8944}	$ \\ \vspace{0.15cm}
	& $	30	$ & $ 	29.5557	$ & $ 	27.0169	$ & $ 	34.4599	$ & $ 	33.7318	$ & $ 	34.6840	$ & $ 	\mathbf{35.5688}	$ \\
	& $	10  $ & $ 	19.9791	$ & $ 	20.2527	$ & $ 	20.1582	$ & $ 	21.8216	$ & $ 	\mathbf{21.9260}	$ & $ 	21.3008	$ \\
Doppler	& $	20$ & $ 	27.3108	$ & $ 	27.6216	$ & $ 	29.3106	$ & $ 	29.5763	$ & $ 	\mathbf{30.8778}	$ & $ 	30.6738	$ \\
	& $	30	$ & $ 	35.7908	$ & $ 	36.0083	$ & $ 	38.5994	$ & $ 	37.2800	$ & $ 	39.9721	$ & $ 	\mathbf{40.0392}	$ \\
	\midrule
\multicolumn{8}{c}{$J_0=7$}\\
\midrule
	& $	10	$ & $ 	13.6838	$ & $ 	17.6491	$ & $ 	16.8629	$ & $ 	\mathbf{18.7340}	$ & $ 	17.1242	$ & $ 	18.4060	$ \\
Blocks	& $	20	$ & $ 	20.4336	$ & $ 	21.4884	$ & $ 	25.1580	$ & $ 	26.0295	$ & $ 	24.7394	$ & $ 	\mathbf{26.3133}	$ \\ \vspace{0.15cm}
	& $	30	$ & $ 	28.6483	$ & $ 	26.2480	$ & $ 	34.0793	$ & $ 	32.6455	$ & $ 	32.7050	$ & $ 	\mathbf{34.6061}	$ \\
	& $	10$ & $ 	12.2659	$ & $ 	13.5925	$ & $ 	16.2727	$ & $ 	\mathbf{18.2628}	$ & $ 	15.4125	$ & $ 	18.1198	$ \\
Bumps	& $	20	$ & $ 	20.0665	$ & $ 	17.7032	$ & $ 	24.9131	$ & $ 	\mathbf{25.7825}	$ & $ 	21.8971	$ & $ 	25.6458	$ \\ \vspace{0.15cm}
	& $	30	$ & $ 	28.5332	$ & $ 	20.9340	$ & $ 	\mathbf{33.9617}	$ & $ 	33.7325	$ & $ 	25.4695	$ & $ 	31.1702	$ \\
	& $	10	$ & $ 	18.1128	$ & $ 	18.7434	$ & $ 	19.9863	$ & $ 	22.1059	$ & $ 	\mathbf{22.4194}	$ & $ 	21.8497	$ \\
Doppler	& $	20	$ & $ 	25.7923	$ & $ 	25.7641	$ & $ 	29.0273	$ & $ 	29.6304	$ & $ 	\mathbf{31.0512}	$ & $ 	31.0435	$ \\
	& $	30	$ & $ 	34.4638	$ & $ 	34.2910	$ & $ 	38.3058	$ & $ 	37.2898	$ & $ 	40.0073	$ & $ 	\mathbf{40.2910}	$ \\
\bottomrule
\end{tabular}
\end{table*}

In addition to SNR measurements,
we also
included RMSE results
shown in Figure~\ref{F:rms}.
The
proposed method
could outperform the competing methods
for a wide range of input SNR
(greater than 10~dB).
It could also surpass
the well-known
\emph{VisuShrink} and \emph{SureShrink}
for all values of input SNR.
For
`blocks' and `Doppler' signals, at lower than 10~dB input SNR,
the
\emph{BayesShrink}
performed slightly better
(less than 3.3\% better).
In the majority of the simulation scenarios,
the \emph{S-median} produced better results than the \emph{VisuShrink} scheme.
This is expected since \emph{S-median} is a
level-dependent version of the \emph{VisuShrink}~\cite{Poorna2008}.

In terms of computational costs,
for the above simulation results,
the \emph{SpcShrink}(1.0\%)
required a number of iterations
ranging from 30 to 46;
whereas
the \emph{SpcShrink}(1.0\%)
required
from~45 to 56~iterations.

We considered a second methodology
for assessing the performance of the wavelet threshold methods.
Now,
we evaluated the SNR measures for each prototype signal
at three different scales
in the multiresolution analysis,
namely: $J_0=3$, $J_0=5$, and $J_0=7$.
We also considered three WGN levels,
1{,}000 Monte Carlo replications,
Daubechies wavelet with eight vanishing moments,
and
signals with $2^{12}$ samples.
Results are presented in Table~\ref{t:results}
in a similar fashion of~\cite{Mert2014}.
The best results are highlighted in bold.

Table~\ref{t:results} shows
a superior performance of the \emph{SpcShrink},
specially with $\alpha_0=1.5\%$.
For $J_0=3$ and $J_0=5$,
we note that the proposed method achieves the best results in almost all cases.
For $J_0=7$,
the results are still good,
keeping the majority of the best figures.
We note that even when the \emph{SpcShrink}
does not achieve the best SNR measures,
it is among the three best results.
In summary,
Table~\ref{t:results}
brings evidence
that the proposed method is
robust even when operating
at different scales ($J_0$)
of wavelet decomposition analysis.

\subsection{Biomedical signal denoising}
\label{s:application}

\begin{figure*}[t]
\centering
\subfigure[Original real IPD]
{\includegraphics[width=0.3\linewidth]{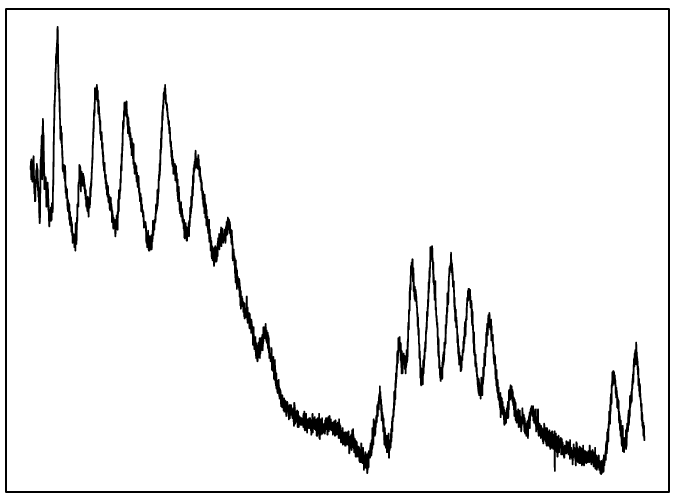} \label{F:ipd}}
\subfigure[Original real ECG]
{\includegraphics[width=0.3\linewidth]{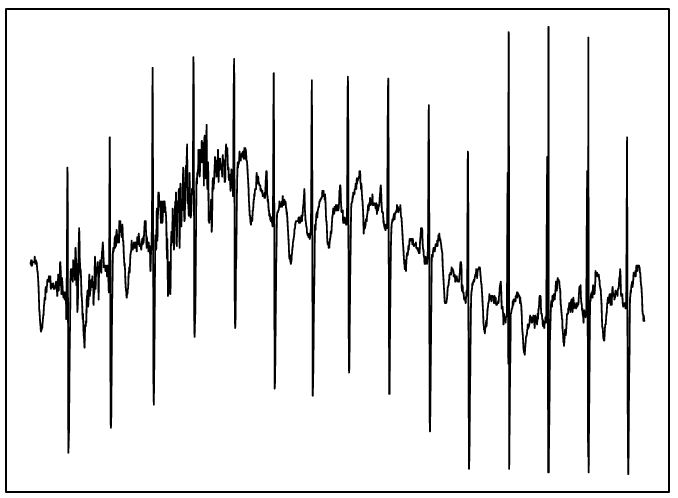} \label{F:ecg}}
\caption{
Biomedical signals considered in denoising experiments.
}
\label{F:bio-signals}
\end{figure*}

In order to illustrate the potential of the proposed \emph{SpcShrink} as a denoising procedure,
we consider two actual biomedical signals, namely:
(i)~an inductance plethysmography data~(IPD),
and
(ii)~an electrocardiogram~(ECG).
Fig.~\ref{F:bio-signals} shows these signals.

\begin{figure*}
\centering

\subfigure[Filtered by the \emph{VisuShrink}]
{\includegraphics[width=0.3\linewidth]{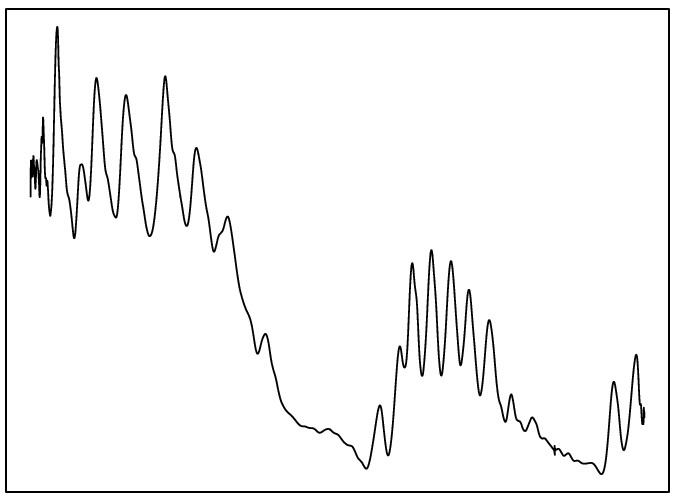} \label{F:ipd_denoised_visu}}
\subfigure[Filtered by the \emph{SureShrink}]
{\includegraphics[width=0.3\linewidth]{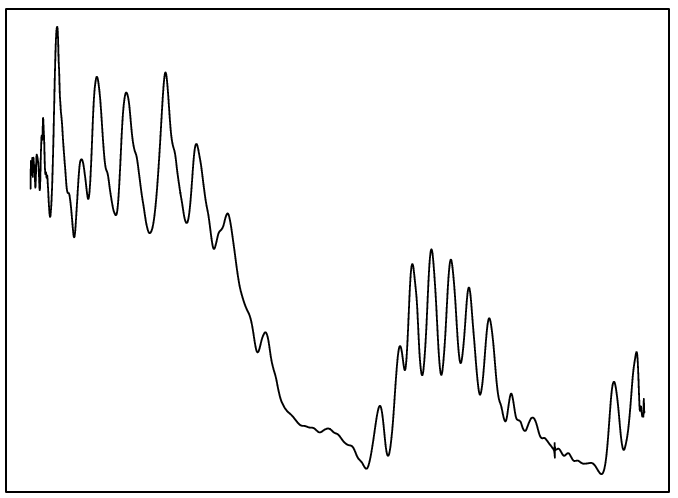} \label{F:ipd_denoised_sure}}
\subfigure[Filtered by the \emph{S-median}]
{\includegraphics[width=0.3\linewidth]{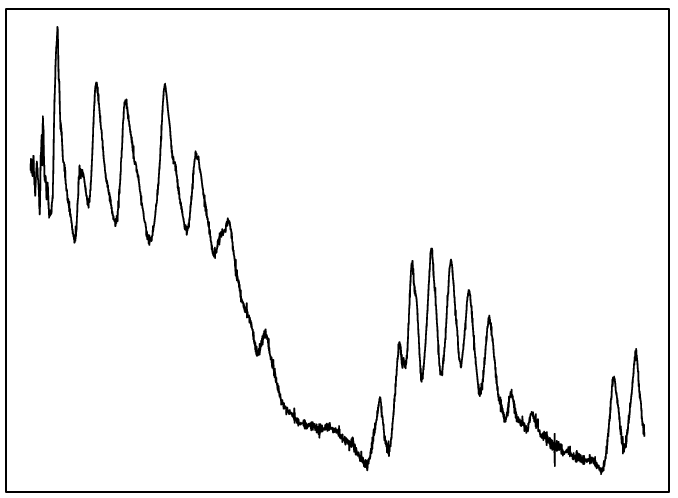} \label{F:ipd_denoised_smed}}
\subfigure[Filtered by the \emph{BayesShrink}]
{\includegraphics[width=0.3\linewidth]{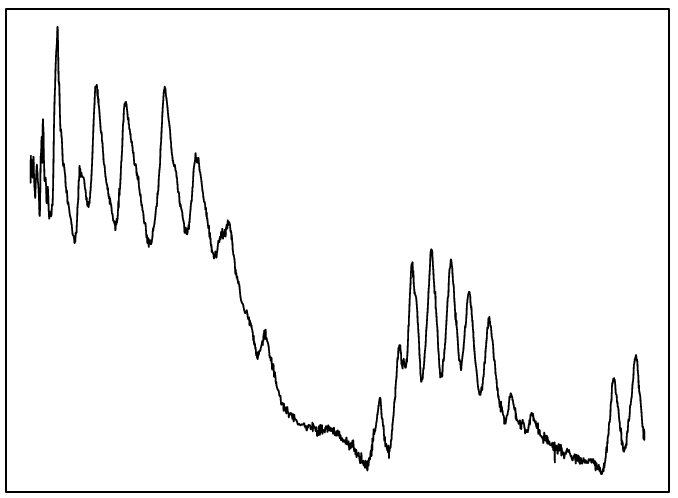} \label{F:ipd_denoised_bayes}}
\subfigure[Filtered by the \emph{SpcShrink}($1.5\%$)]
{\includegraphics[width=0.3\linewidth]{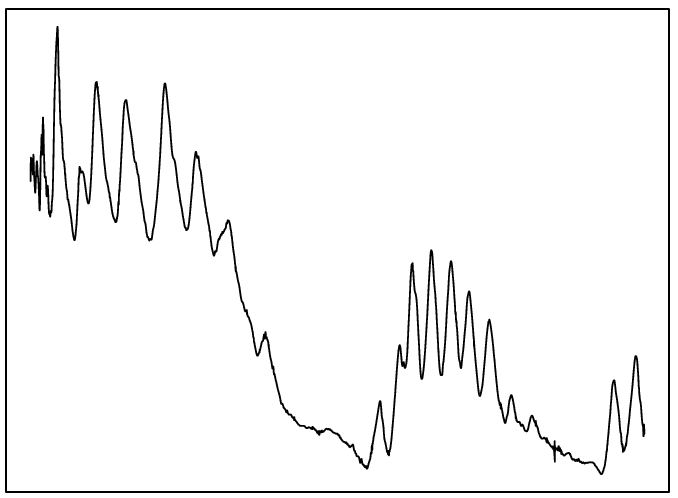} \label{F:ipd_denoised_1.5}}
\subfigure[Filtered by the \emph{SpcShrink}($1.0\%$)]
{\includegraphics[width=0.3\linewidth]{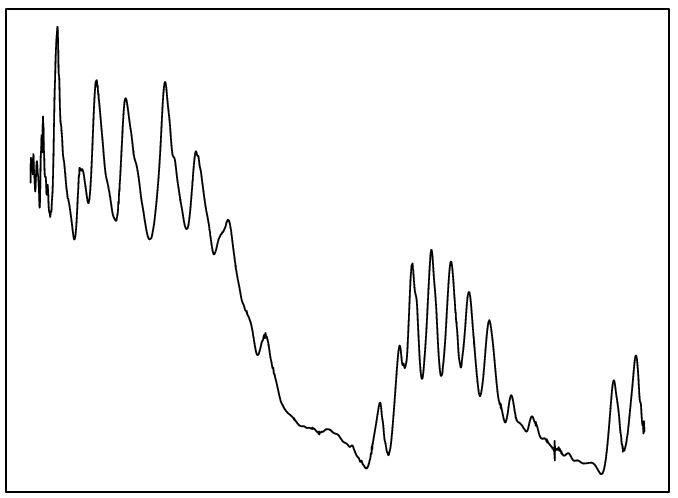} \label{F:ipd_denoised_1}}
\caption{Filtered inductance plethysmography data (IPD) by several methods.}
\label{F:ipd_result}
\end{figure*}

Firstly,
we submitted $2^{12}$ samples of IPD
to the denoising methods.
Such
data was acquired in the context of patient breathing after general anesthesia;
being available in~\cite{wavethresh2013}.
This particular signal was previously described in~\cite{Nason1996}
and was also considered in benchmark tests in~\cite{Johnstone2005}
and \cite{Remenyi2013}.
Figure~\ref{F:ipd_result}
displays
the filtered signal
according to the considered methods.
For these results,
the proposed methods \emph{SpcShrink}(1.0\%) and \emph{SpcShrink}(1.5\%) demanded
only 40 and 56 iterations for convergence,
respectively.
We note that
the proposed algorithm is capable
of removing noise and
preserving the general shape of the signal
as well as its peak intensities and periodicities,
which are significant characteristics to be retained
in a denoised signal~\cite{aldroubi1996}.

\begin{figure*}
\centering
\subfigure[Filtered by the \emph{VisuShrink}]
{\includegraphics[width=0.3\linewidth]{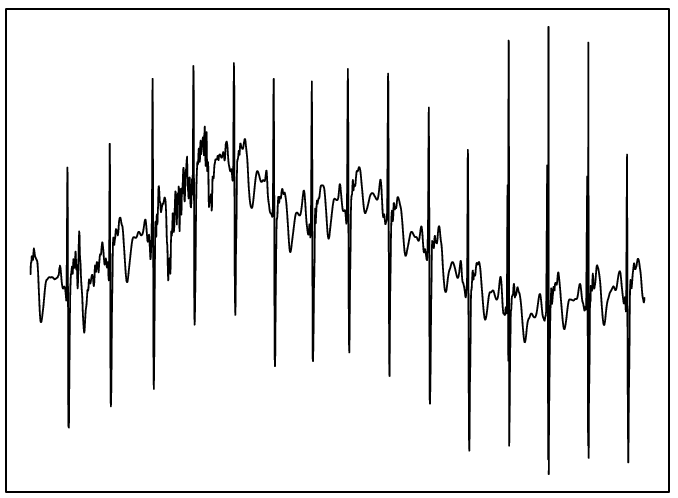} \label{F:ecg_denoised_visu}}
\subfigure[Filtered by the \emph{SureShrink}]
{\includegraphics[width=0.3\linewidth]{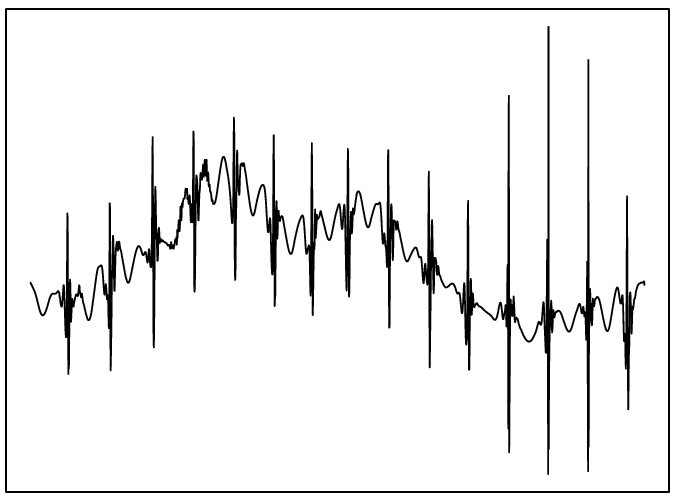} \label{F:ecg_denoised_sure}}
\subfigure[Filtered by the \emph{S-median}]
{\includegraphics[width=0.3\linewidth]{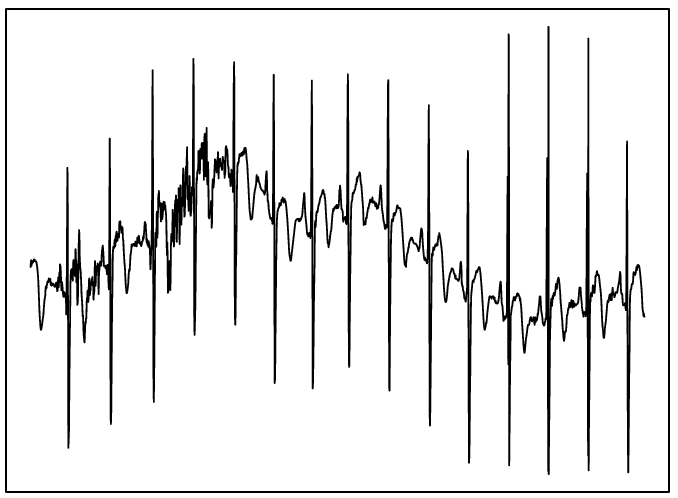} \label{F:ecg_denoised_smed}}
\subfigure[Filtered by the \emph{BayesShrink}]
{\includegraphics[width=0.3\linewidth]{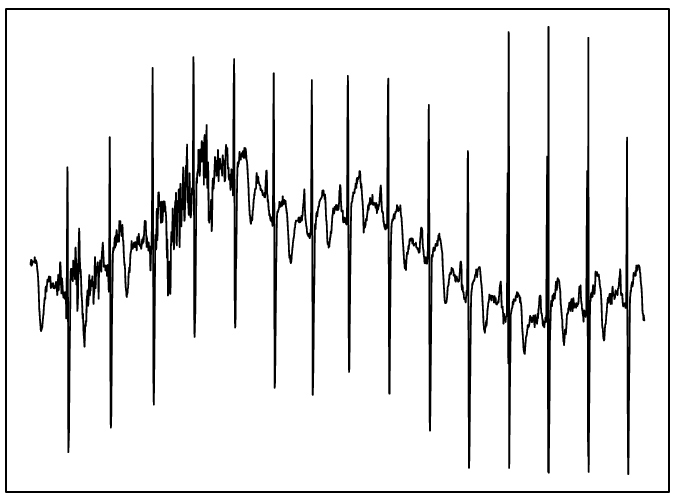} \label{F:ecg_denoised_bayes}}
\subfigure[Filtered by the \emph{SpcShrink}($1.5\%$)]
{\includegraphics[width=0.3\linewidth]{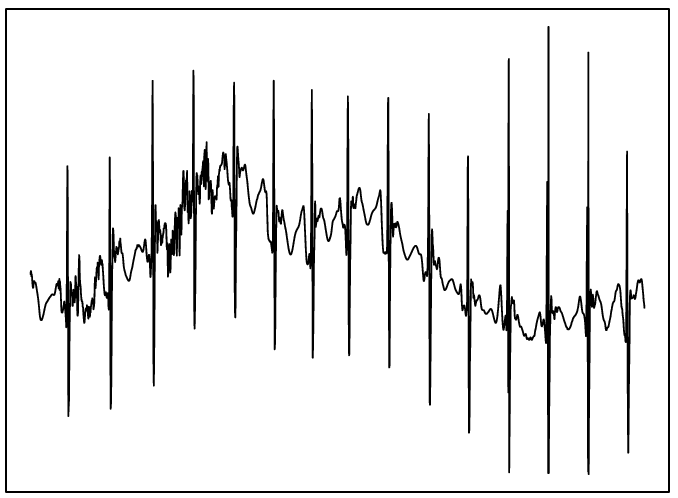} \label{F:ecg_denoised_1.5}}
\subfigure[Filtered by the \emph{SpcShrink}($1.0\%$)]
{\includegraphics[width=0.3\linewidth]{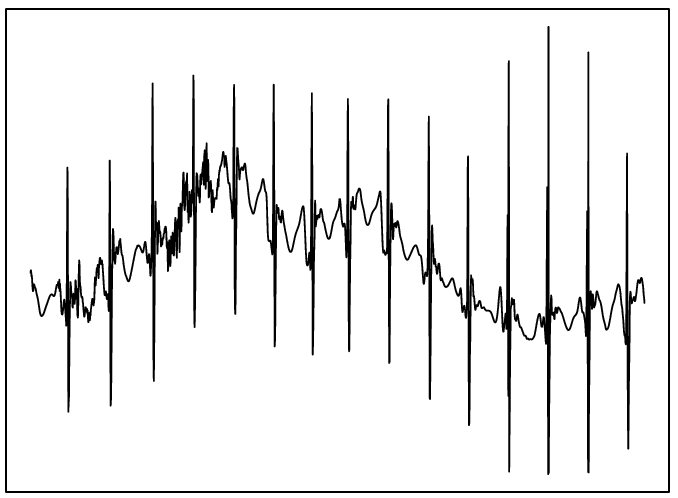} \label{F:ecg_denoised_1}}
\caption{
Filtered electrocardiogram (ECG) signal by several methods.
}
\label{F:ecg_result}
\end{figure*}

Secondly, we considered $2^{11}$ samples of an ECG signal.
Observations sampled from
a patient with arrhythmia
over an interval of 11.37 seconds
at a sampling frequency of 180~Hz.
This data is available in~\cite{R-wavelets}.
For more detailed information regarding this data, see \cite[p.~125]{Percival2000}.
In Fig.~\ref{F:ecg_result} the filtered signals are presented.
In this application,
the \emph{SpcShrink}(1.0\%) and \emph{SpcShrink}(1.5\%) demanded
only 51 and 52 iterations for convergence, respectively.
We note that \emph{SureShrink} flattened the picks of the QRS complex.
The methods \emph{S-median} and \emph{BayesShrink} maintain the most part of the noise in the signal.
Results derived from \emph{VisuShrink} and
the proposed method
are comparable
balancing noise reduction and shape of the ECG signal.

In the discussed biomedical signal processing,
the \emph{SpcShrink} showed similar visual performance
when compared to
the \emph{VisuShrink}.
In ECG filtering the \emph{SureShrink} showed poor results,
whereas it offered good visual quality
for IPD denoising.
Regarding to \emph{S-median} and \emph{BayesShrink} methods,
despite their good performances in the quantitative analysis,
as described in Subsection~\ref{s:monte-carlo},
the resulting signals present visible fine scale roughness.
On the other hand,
the proposed method
excels
both
in quantitative analysis, by maximizing SNR gains
(cf. Section~\ref{s:monte-carlo})
and,
in visual analysis, by effecting smooth signals.
This good performance in both analyses
is due to the strong adaptive trait
present in the proposed method,
which is capable of
discriminating
noise (common cause)
from
signal (special cause)
in an iterative way.

\section{Conclusions}
\label{s:conclusion}

In this work a new signal denoising scheme, called \emph{SpcShrink},
was proposed.
The introduced method was inspired
by the control charts application and based on wavelet shrinkage,
inheriting its own mathematical
justifications from the SPC method.
Iterative search procedures
for establishing the statistical control state
motivated
the search for appropriate threshold values,
which is the main shrinkage step.
The \emph{SpcShrink}
depends on a free parameter $(\alpha_1)$
which allows for
a trade-off between
quantitative error measurements
and
visual quality of filtered signals.
Based on an optimization problem and Monte Carlo simulations results,
we suggest $\alpha_1=1.5\%$ for
SNR gain maximization
and
$\alpha_1 = 1.0\%$
for better visual quality.
The proposed method was
assessed in a Monte Carlo simulation,
with SNR, SNR gain, and RMSE
as figures of merit.
Then it was
compared with several
popular shrinkage schemes,
considering
a variety of test signals and different input noise levels.
In addition to
the good visual performance,
quantitative results
are favorable to the proposed shrinkage method,
which could outperform
\emph{VisuShrink},
\emph{SureShrink},
\emph{BayesShrink},
and
\emph{S-median}
in various scenarios.

\section*{Acknowledgments}

This research was partially supported by FAPERGS, FACEPE, and CNPq, Brazil.

{\small
\singlespacing
\bibliographystyle{siam}
\bibliography{wavelet}
}

\end{document}